\documentclass[10pt,preprint]{aastex}

\slugcomment{Accepted by The Astrophysical Journal}

\def\ltsima{$\; \buildrel < \over \sim \;$}
\def\simlt{\lower.5ex\hbox{\ltsima}}
\def\gtsima{$\; \buildrel > \over \sim \;$}
\def\simgt{\lower.5ex\hbox{\gtsima}}

\received{2001 October 16}
\begin{document}

\title{\bf New Tests of Magnetospheric Accretion in T Tauri Stars}

\author{Christopher M. Johns--Krull\altaffilmark{1} and 
April D. Gafford\altaffilmark{2}}
\affil{Space Sciences Laboratory, University of California, Berkeley, CA
94720,cmj@ssl.berkeley.edu,agafford@ssl.berkeley.edu}
\altaffiltext{1}{Now at Department of Physics and Astronomy, Rice University,
6100 Main St., MS-61, Houston, TX 77005}
\altaffiltext{2}{Also Department of Physics and Astronomy, San Francisco State
University, San Francisco, CA 94132}
\newcommand{\geff}{$g_{\rm eff}$}
\newcommand{\vsini}{\mbox{$v\sin i$}}
\newcommand{\B}{$\vert {\bf B}\vert$}
\newcommand{\kms}{km s$^{-1}$}
\newcommand{\Teff}{$T_{\rm eff}$}
\newcommand{\logg}{$\log\thinspace g$}

\begin{abstract} 

We examine 3 analytic theories of magnetospheric accretion onto classical
T Tauri stars under the assumption that the magnetic field strength does
not vary appreciably from star to star.  From these investigations we
derive predicted relationships among the stellar mass, radius, rotation
period, and disk accretion rate.  Data from 5 studies of the accretion 
parameters of CTTSs are used to test the predicted correlations.  We generally
find that the data do not display the predicted correlations except for
that predicted by the model of Shu et al. as detailed by Ostriker
and Shu and extended here to include non-dipole field topologies. 
Their identification of the trapped flux as an important quantity in the 
model appears to be critical for reconciling the observed data to the theory.
While the data do generally support the extended Ostriker
and Shu predictions, only one of the two studies for which the requisite
data exist show the highest correlation when considering all the relevant
parameters.  This suggests great care must be taken when trying to use
existing observations to test the theory.

\end{abstract}


\section{Introduction} 

Classical T Tauri stars (CTTSs) appear to be roughly solar mass pre-main
sequence stars still surrounded by disks of material which actively
accrete onto the central star.  It is within such disks that solar
systems form.  Understanding the processes through which young stars
interact with and eventually disperse their disks is critical for 
understanding the rotational evolution of stars and the formation of 
planets.  A key question is to understand how young stars can accrete
large amounts of disk material with high specific angular momentum, yet
maintain rotation rates that are observed to be relatively slow
(e.g. Hartmann \& Stauffer 1989, Edwards et al. 1994).  Current models state
that this process proceeds via magnetospheric accretion: stellar magnetic
fields truncate the inner disk and channel accreting material onto the 
stellar surface, preferentially at high latitude.  More generally
in astrophysics, magnetized accretion onto a central source is a common 
theme in various contexts including accretion onto white dwarfs (e.g.
AM Her stars -- Wickramasinghe \& Ferrario 2000), pulsars (pulsating X-ray 
sources -- Ghosh \& Lamb 1979), and possibly black holes at the center 
of active galactic nuclei (e.g. Koide, Shibata, \& Kudoh 1999).  Detailed
tests of magnetospheric accretion theory are necessary for understanding
CTTSs as well as these other classes of objects.

Support for magnetospheric accretion in CTTSs is significant.  Current models
can account for the relatively slow rotation of most CTTSs (Camenzind 1990;
K\"onigl 1991; Shu et al. 1994; Paatz \& Camenzind 1996).  Studies of the
spectroscopic and photometric variability of CTTSs often suggest magnetically
controlled accretion (e.g. Bertout, Basri, \& Bouvier 1988; Bertout et al.
1996; Herbst, Bailer--Jones, \& Mundt 2001; Alencar, Johns--Krull, \& Basri
2001) with the magnetic axis inclined to the rotation axis in some cases
(e.g. Kenyon et al. 1994, Johns--Krull \& Basri 1995, Bouvier et al. 1999).
Models of high resolution Balmer line profiles, computed in the context of
magnetospheric accretion, reproduce many aspects of observed line profiles
(Muzerolle, Calvet, \& Hartmann 1998, Muzerolle et al. 2000).  T Tauri stars
are observed to be strong X-ray sources indicating the presence of strong 
magnetic fields on their surfaces (see Feigelson \& Montmerle 1999 for a 
review), and CTTSs are
observed to have strong surface magnetic fields (Guenther et al. 1999; 
Johns--Krull et al. 1999b; Johns--Krull et al. 2001), and strong magnetic 
fields have been observed in the formation region of the \ion{He}{1} emission
line at 5876 \AA\ (Johns--Krull et al. 1999a; Johns--Krull \& Valenti 2000), 
which is believed to be produced in a shock near the stellar surface as the 
disk material impacts the star.

Despite these successes, open issues remain.  Current theoretical models
assume a dipole geometry for the stellar magnetic field; however, Johns--Krull
et al. (1999a) and Johns--Krull and Valenti (2001) show that the surface
fields on BP Tau and TW Hya are not dipolar.  On the other hand, it is
expected that the dipole component of the field should dominate at distance
from the star where the interaction with the disk is taking place, so this 
may not contradict current theory.  Even in the case of the complex magnetic
topology of the Sun, the dipole component appears to become dominant at
$2.5 R_\odot$ or closer (e.g. Luhmann et al. 1998).  For expected disk
truncation radii of 3 -- 6 $R_*$ in CTTSs, this suggests the dipole component
will govern the stellar interaction with the disk.  On the other hand,
Stassun et al. (1999) find no correlation
between rotation period and IR signatures of circumstellar disks in a sample
of 254 stars in Orion, leading them to question the validity of magnetically
mediated disk locking which is intrinsic to magnetosphereic accretion
theory.  Bertout and Mennessier (1996) review points in favor of both
the magnetospheric accretion model and the original axisymmetric, equatorial
boundary layer model (based on the work of Lynden--Bell \& Pringle 1974) 
which came before.  While the magnetospheric model remains the favorite
in the current literature, Bertout and Mennessier (1996) conclude that the
existing data does not favor either model.  Thus, the overall success of
magnetospheric accretion as the best description of accretion onto CTTSs
remains in some doubt.

The current equilibrium theories of magnetospheric accretion predict
specific relationships between the stellar mass ($M_*$), radius ($R_*$),
rotation period ($P_{rot}$), magnetic field ($B_*$), and mass accretion
rate ($\dot M$) (e.g. Johns--Krull et al. 1999b).  Unfortunately, 
observation of all these quantities are only available for a handful of 
stars, with magnetic field measurements the most lacking.  One could look
for correlations among the other quantities to test the theories.  Ghosh
(1995) investigated the rotation and accretion properties of a number of
T Tauri stars (TTSs) by assuming a dynamo relation for TTS magnetic fields 
in terms of Rossby number (convective turnover time divided by rotation 
period) since there were almost no magnetic field measurements for TTSs
available at
the time.  Ghosh (1995) found that TTSs generally cluster below the limiting
rotation rate as is to be expected; however, no clear correlations were
seen in the data.  Johns--Krull, Valenti, and Linsky (2001) have shown
that TTS Rossby numbers generally place them in the region of saturated
magnetic activity compared to main sequence and evolved stars, so it is
not clear that parameterizing the stellar field in terms of the Rossby 
number is appropriate for these stars.  Furthermore, TTSs do not show a large
range of magnetic field properties as described below.  While we are 
unaware of any other systematic studies along this line, Muzerolle, Calvet,
and Hartmann (2001) note that current theory predicts a correlation between
rotation period and mass accretion rate which has not been observed.
Muzerolle et al. (2001) suggest that variations in the stellar magnetic field
strength from star to star may account for the lack of correlation.

Johns--Krull et al. (2001) show that the mean photospheric magnetic fields 
of 6 TTSs range only between 2.1 and 2.7 kG, suggesting a surprising 
uniformity to the field strengths on these stars.  The field on T Tau 
obtained by Guenther et al. (1999) also falls in this range.  These 
observations are of the magnetic field averaged over the entire stellar 
surface, while for the current discussion we are most interested in the 
field in the actual accretion zones.  Johns--Krull et al. (1999a) discovered
a magnetic field in the \ion{He}{1} 5876\AA\ emission line on BP Tau from 
observations of circular polarization, arguing that the narrow component of
this line likely forms in the postshock region as material accretes onto the
stellar surface (see also Hartmann, Hewett, \& Calvet 1994 and Lamzin 1995).
The field detected is 2.5 kG, which is a lower
limit to the true field due to the unknown angle between the field lines
and the line of sight.  Johns--Krull et al. (2001) also present similar
spectropolarimetric observations of 4 CTTSs over 6 nights, finding 
fields in the \ion{He}{1} line formation region that vary smoothly with
time, probably due to stellar rotation which suggests the fields
participating in the accretion flow are stable at least on a rotational
time scale (see also Valenti, Johns--Krull, \& Hatzes 2001).  In all cases,
the field determined from the \ion{He}{1} polarization is equal to or
lower than the mean photospheric field, as is expected if the fields are
in fact the same but the \ion{He}{1} measurement is reduced as a result of
field line inclination to the line of sight.

As mentioned above, the theoretical relationships predicted by magnetospheric 
accretion theory involve many parameters, not just rotation period and mass
accretion rate.  Since all the exisiting observational data suggests relative 
uniformity of surface magnetic field strengths on CTTSs (at least in Taurus
where the most detailed studies have taken place), we have examined 
several sources of data from the literature to look for the expected 
correlations assuming the field strength is in fact constant from 
star to star.  This constancy of the magnetic field may represent some
limit to the dynamo operating in these young stars.  Certainly, the details
of such a dynamo are far from well understood; however, Johns--Krull et al.
(2001) show that within the context of other active stars, TTS are expected
to show saturated magnetic activity based on their rotation and convective
parameters, perhaps giving further justification to assuming a constant 
magnetic field.  In \S 2 we review the predicted relationships between the 
various stellar and accretion parameters.  In \S 3 we examine literature 
compilations of the requisite parameters for testing the predicted 
correlations.  Section 4 presents the analysis of this data.  In \S 5 we 
discuss our results, and our conclusions are given in \S 6.

\section{Magnetospheric Accretion Predictions}

Johns--Krull et al. (1999b) examined three analytic theories (K\"onigl 1991,
Cameron \& Campbell 1993, Shu et al. 1994) of magnetospheric accretion,
presenting equations for the stellar magnetic field strength as a function
of $M_*$, $R_*$, $P_{rot}$, and $\dot M$.  These studies are outgrowths of
the pioneering work done by Ghosh and Lamb (1979) on accreting neutron
stars.  The work of K\"onigl (1991) and Cameron and Campbell (1993) are
very closely related to that of Ghosh and Lamb (1979) in that these papers
only consider accretion onto the star.  The Shu et al. (1994) work is different
in that it is an integrated theory of accretion onto the central star plus
a wind which is driven off the disk at the truncation point (the X-region).
While coefficients representing such things as the efficiency of magnetic 
coupling to the disk vary from one study to the other, the predictions from
K\"onigl (1991) and Shu et al. (1994) depended in the same way on the above 
stellar and accretion parameters for an assumed dipolar magnetic field.  
Holding the stellar magnetic field constant, K\"onigl (1991) and Shu et al.
(1994) predict that
\begin{equation}
R_*^3 \propto M_*^{5/6} \dot M^{1/2} P_{rot}^{7/6}.
\end{equation}
In the case of Cameron and Campbell (1993), the relationship is
\begin{equation}
R_*^3 \propto M_*^{2/3} \dot M^{23/40} P_{rot}^{29/24},
\end{equation}
which is very similar in its dependencies to K\"onigl (1991) and Shu et al.
(1994).  In fact, scatter plots created using equation (2) are virtually
identical to those created using equation (1).  Therefore, we only show plots
based on equation (1); however, the results for equation (2) are cited in
the tables and discussed in the text.

In a more detailed analysis of the magnetospheric accretion flow itself,
Ostriker and Shu (1995) identified the importance of the notion of trapped
flux in the Shu et al. (1994) theory.  While still assuming a dipole
stellar field, Ostriker and Shu noted that what their theory really 
predicted was the amount of magnetic flux trapped in the X-region.  In their
preferred model, this is 1.5 times the magnetic flux which would thread the
disk exterior to the truncation point (the co-rotation radius in their model)
in an undisturbed dipole field.  For the system to remain in equilibrium,
this is the flux that must be trapped in the X-region independent of the
topology of the magnetic field on the stellar surface.  Armed with this
information, we can derive a new relationship among the observable 
parameters which is unique to the Ostriker and Shu (1995) treatment of
the Shu et al. (1994) theory.

Re-writing equation (4.1) of Ostriker and Shu (1995) for the radius of the
X-point, $R_x$, we have
\begin{equation}
R_x = \Phi_{dx}^{-4/7} \Bigl( {\mu_*^4 \over G M_* \dot M_D^2} \Bigr)^{1/7}
\end{equation}
where $\Phi_{dx}^{-4/7}$ is a constant evaluated by Ostriker and Shu (1995)
to be 0.923 in their preferred model, $\mu_*$ is the dipole moment of the star,
$G$ the gravitational constant, and $\dot M_D$ is the mass accretion rate
through the disk.  The general form of this equation is well known from
many studies of accretion onto compact objects (see for example White \&
Stella 1988).  In the Shu et al. (1994) theory, the disk mass accretion
rate is not equal to the accretion rate onto the star (the quantity measured 
in the observations described below) since the theory also
incorporates a magnetocentrifugally driven wind from the X-region; however,
the two are related by a factor of order 2/3.  

For a purely dipolar field, the magnetic flux threading the disk exterior
to the point $R_x$ is
\begin{equation}
\phi_x = {2 \pi \mu_* \over R_x}.
\end{equation}
In the preferred model of Ostriker and Shu (1995) the total flux trapped 
in the X-region is $\phi_t = 1.5 \phi_x$, of which a third participates in
the accretion funnel flow of material onto the star (another third is in the 
dead zone, and the final third is part of the wind).  Thus, the magnetic flux
threading the X-region which directs disk material onto the star is
$0.5 \phi_x$ independent of the exact topology of the field at the stellar
surface, as long as the dipole dominates in the X-region.  Equating
this magnetic flux to that on the star in the accretion zones, we have
\begin{equation}
B_* A_* = 0.5 \phi_x = {\pi \mu_* \over R_x}
\end{equation}
where $A_*$ is the area on the surface of the star on which the accreting
disk material falls.

If we recall that $R_x$ in the Shu et al. (1994) theory is the co-rotation
radius, where the Keplerian angular velocity in the disk is equal to the
angular velocity of stellar rotation, we can solve equation (5) for $\mu_*$,
substitute the result into equation (3) and get 
\begin{eqnarray}
B_* A_* & = & 2^{-1/2} \pi^{1/2} G^{1/2} \Phi_{dx} M_*^{1/2} 
\dot M_D^{1/2} P_{rot}^{1/2}
\nonumber\\
& = & 8.88 \times 10^{24} \Bigl({M_* \over M_\odot}\Bigr)^{1/2}
\Bigl({\dot M_D \over 10^{-7} M_\odot\,\, {\rm yr^{-1}}}\Bigr)^{1/2}
\Bigl({P_{rot} \over 1\, {\rm day}}\Bigr)^{1/2} \,\,\, {\rm G\ cm^2}
\end{eqnarray}
after some manipulation.  In this paper, we assume $B_*$ does not vary from
star to star and $A_* = 4 \pi R_*^2 f_{acc}$ where $f_{acc}$ is the 
filling factor of accretion zones, so equation (6) implies
\begin{equation}
R_*^2 f_{acc} \propto M_*^{1/2} \dot M^{1/2} P_{rot}^{1/2}
\end{equation}
as a unique prediction of the Shu et al. (1994) theory as detailed by
Ostriker and Shu (1995) and expanded here to incorporate a non-dipole field
topology near the stellar surface.  In equation (7), $\dot M$ is the mass
accretion rate onto the star and $\dot M = \alpha\dot M_D$ where $\alpha \sim
2/3$ since the wind carries away about one third of the {\it disk} accretion
rate (Shu et al. 1994).   Hartigan, Edwards, and Ghandour (1995) attempted
to observationally determine the ratio of mass loss to mass accretion
rate, finding a value of 0.01; however, more recent estimates of this
ratio are $\sim 0.1$ (see review of K\"onigl \& Pudritz 2000) so the
Shu et al. (1994) value seems quite reasonable given the current state
of the observations.

In equation (7) above, the filling factor, $f_{acc}$, is the filling factor
of the accretion spots on the stellar surface, which is not the same as
the filling factor of the magnetic field.  In the theories which use a strict
dipole field geometry in which the magnetic axis is aligned with the rotation
axis (the typical assumption), the magnetic filling factor is 1.0, but 
accretion onto the star occurs only in axisymmetric rings located at high 
latitude.  Thus the filling factor of accretion zones is expected to be
significantly less than 1.0, and its value is strongly influenced by the small
scale topology of the magnetic field at the stellar surface.  On the other 
hand, it is the bright accretion shock emission produced in these regions 
which is believed to give rise to the UV continuum excess and optical veiling
emission seen in CTTSs (e.g. Calvet \& Gullbring 1998).  As a result, it is
possible to observationally determine $f_{acc}$, and this has been done for 
reasonably large samples of stars by Valenti, Basri, and Johns (1993) and 
Calvet and Gullbring (1998).  As discussed further below, accurate estimates
of $f_{acc}$ are required to test the correlations predicted by equation (7),
and these estimates can be difficult to make.

In this paper we focus on the above mentioned theories since they are largely
analytic and permit the derivation of relatively simple expressions which
show the dependence of stellar and accretion parameters on one another
in the equilibrium situation.  These expressions can be tested in a
rather straightforward way using existing observational data.  This is
not to suggest that the above studies are the only such studies of
magnetospheric accretion in CTTSs.  There are a number of others, some
of which derive their results from numerical studies of the governing
equations (e.g. Paatz \& Camenzind 1996), while some focus on the 
instabilities that can result between the star, its magnetosphere, and
the disk which may launch winds and jets (e.g. Hayashi, Shibata, \&
Matsumoto 1996; Goodson, Winglee, \& B\"ohm 1997; Miller \& Stone 1997;
Goodson, B\"ohm, \& Winglee 1999; Ferreira, Pelletier, \& Appl 2000).  These
studies are not directly testable in the fashion explored in this paper;
however, important tests of many of these works may be possible with
variability studies (see Goodson et al. 1999).  Finally, we note that
CTTSs are variable as has been mentioned in \S 1.  This variability is not 
accounted for in the equilibrium theories examined in this paper, but
is not necessarily evidence that these theories should be abandoned.  The
theories may be perfectly adequate to describe the average conditions of 
CTTSs.  In this case, we expect to see the predicted correlations in the data,
and the variability (as well as measurement error) should produce scatter 
about the mean relations.  It is these correlations we are looking for here.

\section{Data Used}

The goal of this paper is test the proportionalities predicted in equations
(1), (2), and (7) by using data from the literature to look for the 
predicted correlations.  For consistency within each of the studies we
employ, we use the stellar and accretion parameters from each individual
study when testing for the predicted correlations instead of trying to
come up with a single best estimate for each of the relevant parameters.
The rotation periods are all taken from photometric monitoring studies which
do not depend on these stellar parameters, so we apply the detected periods 
to all the accretion data sets.

Mass accretion rates onto large samples of CTTSs have been estimated by 
Valenti et al. (1993 -- hereafter VBJ); Hartigan, Edwards, and Ghandour 
(1995 -- hereafter HEG); Gullbring et al. (1998 -- hereafter GHBC); Calvet 
and Gullbring (1998 -- hereafter CG98); and White and Ghez (2001 -- hereafter
WG01).  Of these papers, only VBJ and CG98 estimate the filling factor, 
$f_{acc}$, of accreting material on the stellar surface, which is required 
to test the relationship given in equation (7).  CG98 do not separately 
estimate, $T_{eff}$, $M_*$, and $R_*$, instead using the values tabulated by 
GHBC which are reproduced in the tables here.  For the 4 continuum stars (CW 
Tau, DG Tau, DL Tau, and DR Tau) from CG98 which are used in this study, CG98
{\it assume} $R_* = 2 R_\odot$ and $M_* = 0.5 M_\odot$, which we use here.  In
addition, CG98 do not tabulate $\dot M$ specifically, but the mass loss rate
can be calculated from their equation (11) and the quantities tabulated in 
the paper.  We note that in CG98, they do not perform a detailed fit to the
data presented in GHBC, instead adopting the values from their grid of models 
which gives the best fit to the data as estimated by eye.  However, this 
should be adequate for our purpose of looking for general correlations among
different parameters.  Rotation period data comes from the following papers in
order of preference: Bouvier et al. 1995, Bouvier 1990, Bertout et al. 1996,
and Bouvier et al.  1993.  Tables 1 through 5 give the final compilation
of stars and their relevant parameters for the 5 studies examined here.
In a few cases (noted in the tables), required quantities such as the 
stellar radius are not tabulated in the original studies, but can be estimated
from the tabulated stellar luminosity and effective temperature.  For those 
studies which only tabulate spectral type, the spectral type -- effective 
temperature calibration of Johnson (1966) is used in this paper to estimate
$T_{eff}$.

Some of the stars used in this study are now known to be binary, while this
was not known to be the case in earlier studies.  In addition, pre-main
sequence evolutionary tracks have been refined over time, which influences
the derivation of stellar parameters.  For many
of the binaries, their separation is large enough (DK Tau, GK Tau, RW Aur)
that they do not fall in the slit used by HEG, while WG01 explicitly take
into account the binary nature of these stars.  For the studies based on
low to medium resolution spectroscopy (VBJ, GHBC, CG98), the slits used
were wide enough to encompass both members of the binary systems.  However,
it is generally the case (DF Tau, DK Tau, GG Tau A, RW Aur) that the primary
star dominates the optical and short wavelength light (Mathieu 1994; Ghez,
White, \& Simon 1997; Duchene et al. 1999; WG01) from which the accretion 
parameters are derived.  As a result, the accretion rates and filling factors
apply to the primary in the case of those stars with originally unknown 
companions.  Inclusion of the WG01 results in this study will also serve to
see if accurate accounting for binarity significantly changes the results.
The use of different pre-main sequence tracks and differences in the 
extinction corrections used (as well as unknown binaries) between the 
different studies results in small (generally less than a factor of 2) 
changes in the 
stellar mass and radius from one study to the other, but these differences 
are small compared to the full range of parameters explored in the samples.
We choose to use the samples as is since changes made in one parameter can 
also affect others (such as the mass accretion rate), so here we hope to 
maintain consistent sets of parameters from one study to the next as we 
look for general correlations among the parameters.

In addition to observational issues related to the influence of binaries,
it is fair to consider the theoretical influence of binary interactions.
This paper is concerned with the star-disk interaction, and the theories 
examined here have the disk accretion rate as a free parameter.  Therefore,
if CTTSs in binaries have different disk accretion rates compared to single 
CTTSs, this difference should naturally be accounted for.  Theoretical 
effects studied in CTTS binary systems inlclude disk truncation and 
circumbinary replenishment (e.g. Artymowicz \& Lubow 1994, 1996) and changes 
to disk accretion rates (e.g. Ostriker, Shu, \& Adams 1992), but most authors 
consider the star-disk interaction to be the same as for single CTTSs
(e.g. Armitage, Clarke, \& Tout 1999).  Observationally, WG01 find no 
difference in the accretion rate as a function of binary separation 
in the range 10 - 1000 AU, and Simon \& Prato (1995) find no difference in 
disk lifetime between binary and single CTTSs.  Thus, it is unlikely there
is any real difference in the star-disk interaction for CTTSs in relatively
wide binaries such as those considered here compared to single CTTSs.  As
a result, we include both single CTTSs and CTTSs in binaries in the current
analysis, but we separate them by plot symbol for clarity.  Stars identified
as binary in WG01, Mathieu (1994), or Ghez, Neugebauer, and Matthews (1993)
are labelled as binary; while stars indicated as single in WG01 are 
labelled so here.

As mentioned above, VBJ and CG98 are the only studies which determine 
the filling factor, a necessary quantity for testing equation (7).  
Generally speaking, the mass accretion rate is
\begin{equation}
\dot M = 4 \pi R_*^2 f_{acc} \rho v,
\end{equation}
where $\rho$ is the density of the infalling material and $v$ is the 
velocity at which it is coming in.  Therefore, 
\begin{equation}
f_{acc} = {\dot M \over 4 \pi R_*^2 \rho v}.
\end{equation}
HEG, GHBC, and WG01 all give $\dot M$ and either $R_*$ or the data 
necessary to derive it.  One could either simply take single representative
values for $\rho$ and $v$ or try to estimate them individually for each 
star.  In either case though, the value of $f_{acc}$ determined depends 
explicitly on the mass accretion rate.  Therefore, one would naturally
expect a good correlation between the left and right side of equation (7)
since $\dot M$ appears on both sides.  We have estimated $f_{acc}$ from 
equation (9) and do indeed find good correlations for equation (7), but this
cannot be viewed as support for our extended version of the Ostriker and Shu
(1995) model, so we do not include these plots for the data of HEG, GHBC, or
WG01.

In summary then, we use data from 5 observational studies to look for the
correlations predicted in equations (1), (2), and (7).  Of these, 3 studies
do not independently determine the filling factor of accretion spots, 
$f_{acc}$, and therefore are only suitable to test the relationships of the
simple (dipole) accretion theories embodied in equation (1) and (2).  These
three studies are HEG, GHBC, and WG01.  The investigations of VBJ and CG98 do
independently estimate $f_{acc}$ and are therefore suitable for testing the
prediction of our modified Ostriker and Shu (1995) model, as well as the
predictions of the simple models.  Thus, VBJ and CG98 can be used to test
equations (1), (2), and (7).

\section{Analysis}

\subsection{Simple Accretion Theories}

Figures 1a, 2, 3, 4a, and 5 show the results for the data from the 5 studies
when looking for the correlation predicted by equation (1), which results
from assuming a dipole magnetic field geometry with a constant field strength
from star to star following K\"onigl (1991), Shu et al. (1994 -- and the 
unmodified Ostriker and Shu 1995 theory).  In each figure, the quantity
on the right hand side of the equation (1) is plotted versus the quantity on
the left hand side of the equation.  For simplicity in the plot units, we have
expressed the stellar mass and radius in solar units for each star, and the 
mass accretion rate in units of $10^{-7}$ M$_\odot$ yr$^{-1}$.  As mentioned 
above, the scatter plots which result from equation (2) following Cameron
and Campbell (1993) look almost identical
to those shown in these figures and are therefore not shown separately.  
Filled circle plot symbols denote single CTTSs, and asterisks denote CTTSs 
in binary systems.  In all the figures, the expected correlation is shown 
with a dashed line.  For those plots which do show a significant correlation 
(see below), the best fit line is shown in a solid line.  In general, the 
expected relationship for equation (1) is only weakly present, with the data
of HEG showing the strongest correlation.

We quantify these results by computing the linear correlation coefficient
(also known as Pearson's $r$) and its associated false alarm probability,
$P_f$ (Press et al. 1986).  For each investigation, the associated 
correlation coefficient and false alarm probability are computed by using
the logarithm of the quantities on each axis of the plots, thus giving the
linear correlation of the scatter plots actually shown in these log-log 
plots.  This is done to prevent the calculation of the correlation from 
being overly influenced by one or a few points with very large
values on either axes.  The resulting correlation coefficients and false
alarm probabilities are given in Table \ref{corr}.  It is readily apparent
from this table that when testing the correlations predicted by equations
(1) or (2), only the HEG data produces a false alarm probability $\leq 0.01$,
which is typically indicative of a real correlation.  The data of HEG give
$P_f = 0.009$ in the case of equation (2), suggestive of a significant 
correlation.  The best fit line (solid line in the plot) is determined from a
standard least squares analysis, assuming each data point has the same
uncertainty in the y-axis quantity.  To estimate the uncertainty in
the slope (predicted to be 1 by the theory), we then use the best fit line
to estimate observed residuals for each point.  The standard deviation of
these residuals is then assumed to be the $1\sigma$ uncertainty of each point
in the y-axis quantity, which in turn gives an uncertainty for the slope in
the least squares fit.  Given the number of potential uncertainties that
go into estimating the error bars for the points in each figure, we feel
this means of estimating the final uncertainty of the slope from the 
scatter plots is the least biassed.  The slope and its uncertainty are given
in Table \ref{corr} for that equation [(1), (2), or (7) discussed below]
which produces the best correlation for each data set.  For the case of
HEG data and equation (2), the slope is $0.88 \pm 0.29$ which is 
insignificantly different from the expected value of 1.0.

\subsection{Modified Ostriker and Shu}

\subsubsection{Initial Tests}

Figures 1b and 4b show the results for the data from VBJ and CG98 respectively
when looking for the correlation predicted by equation (7), which results
from using the trapped flux concept of Ostriker and Shu (1995) while
abandoning a pure dipole magnetic field geometry and assuming a constant
stellar magnetic field strength.  The plot symbols have the same meaning as
given above, as do the solid and dashed lines.  Both of these plots show
much stronger correlations than present in the data when testing equations
(1) and (2).  This is also apparent in the correlation coefficients and
false alam probabilities given Table \ref{corr}.  Also given in the table
is the best fit slope and uncertainty calculated in the same way as described
above.  The expected slope is 1.0 and the VBJ data give this slope to within
the uncertainties.  The CG98 data give a slope substantially below this 
expected value.  This low slope is largely driven by the two stars with the
lowest values of $R_*^2 f_{acc}$; however, excluding them still gives a slope
which is too low at the $4\sigma$ level.  This gives some concern, but as
mentioned before (and discussed further below), CG98 do not actually fit
their models to the observations, indicating that their study may not be
the best suited for strictly testing the predicted correlations.  Nevertheless,
it is certainly true that equation (7) produces far better correlations in
the data than either equations (1) or (2).

\subsubsection{A Closer Look at VBJ and CG98}

The data from both VBJ and CG98 appears to support the Ostriker and Shu
(1995) theory as extended here to include non-dipole field topologies,
while the data from all 5 observational studies at best only
weakly supports the other magnetospheric accretion theories.  As a
result, it is important to look closely at VBJ and CG98 to see just how well
they agree with equation (7).  We also note from equation (8) that
generally, $\dot M$ depends on $f_{acc} R_*^2$, so there should be some 
correlation between these quantities.  The observations primarily
determine the accretion luminosity, $L_{acc}$, which is then used to
determine the mass accretion rate either directly by applying an
equation of the form 
\begin{equation}
\dot M = \beta R_* L_{acc} / G M_*
\end{equation}
where $\beta$
depends on whether accretion occurs through an equatorial boundary layer
or through a magnetospeher; or a model is used in the case of CG98 to fit
the accretion luminosity and the resulting parameters give the mass
accretion rate.  At the same time, the filling factor, $f_{acc}$, is usually
determined from the equation 
\begin{equation}
L_{acc} = F_{acc} 4 \pi R_*^2 f_{acc}
\end{equation}
where $F_{acc}$ is the emission per unit area from the accretion zones and
is determined by fitting the observations to a model, either a hydrogen
slab model in the case of VBJ or a shock model for CG98.  Thus, both
the determination of $\dot M$ and $f_{acc}$ primarily depend on $L_{acc}$
and the value of $R_*$.  Therefore, it is important to see how well 
$\dot M$ is correlated with $R_*$ (or $R_*^2$, but when looking at the
log of the quantities, it does not matter which power is considered
when evaluating the correlation between $\dot M$ and $R_*$) and $f_{acc}$ 
separately, as well as with the product $R_*^2 f_{acc}$ to see if the 
correlations seen in Figures 1b and 4b are driven by these potential 
correlations.  Indeed, CG98 point out that $f_{acc}$ appears to be well 
correlated with $\dot M$ in their study, making this a real concern.  On the
other hand, we already know that $\dot M$ is not too well correlated with
$R_*$ since the top two panels in each figure do not show strong correlations.

In table \ref{closer} we give the correlation coefficients and resulting 
false alarm probabilities when comparing log($\dot M^{1/2}$) to log($f_{acc}$), 
log($R_*^2$) and log($f_{acc} R_*^2$) using the data from VBJ and CG98.  These
correlations are to be compared with the values in the last two columns
of Table \ref{corr} which gives the correlation coefficient and false
alarm probability when comparing all the parameters relevant to our extended
Ostriker an Shu theory [log($M_*^{1/2} \dot M^{1/2} P_{rot}^{1/2}$) versus
log($f_{acc} R_*^2$)].  In the case of VBJ, there is no correlation between
the mass accretion rate and the filling factor, some between $\dot M$ and
$R_*^2$ and a real correlation between $\dot M$ and $f_{acc} R_*^2$, but the
full Ostriker and Shu (1995) comparison (Table \ref{corr}) produces the
best correlation.  Thus, it appears that the inclusion of the rotation
period and the mass do improve the results as predicted by the theory.
The case is not quite the same for CG98, with the $\dot M$ showing the
best correlation with $f_{acc} R_*^2$, but the correlation based on equation
(7) is nearly as high.  It is certainly true that the
CG98 shock model is the most physical of the two studies; however, CG98 do
not actually fit the observed data, but pick by the eye the closest model
from their grid to match the observations.  It is also apparent from their
Figure 6 that they do not generally match the observations (particularly
the slope of the Paschen continuum) as well as VBJ, so it cannot be assumed
that their values for various accretion parameters are more accurate. 

We can look further at the results of VBJ.  We substitute equation (10) into
equation (7) to eliminate $\dot M$ and then use equation (11) to eliminate
$f_{acc}$.  Dropping constants and re-arranging, we then find
\begin{equation}
R_*^{-1/2} L_{acc}^{1/2} \propto F_{acc} P_{rot}^{1/2}.
\end{equation}
In the case of VBJ, $F_{acc}$ is determined by a model fit and its value
is largely driven by the shape of the Paschen continuum and the size of 
the Balmer jump.  Equation (12) above involves parameters which are
independently determined to the greatest extent possible within the context
of the VBJ study.  The accretion luminosity is given by VBJ and we 
reproduce these in Table \ref{vbjt} for each star.  Since we have access to
the models used by VBJ, we compute $F_{acc}$ by integrating over wavelength
the excess flux produced by the best fitting hydrogen slab model for each 
star and also give these values in Table \ref{vbjt}.  We note that VBJ do
not estimate $f_{acc}$ using equation (11).  Use of this equation with
values of $F_{acc}$ from Table \ref{vbjt} to estimate $f_{acc}$ produce
values which are systematically larger that those given by VBJ.  The model 
of VBJ used the observations to determine the filling factor of the 
visible hemisphere ($\pi R_*^2$) covered by the boundary layer (accretion
zones).  The systematic difference between the two estimates of $f_{acc}$
reflect the unknown forshortening of the emitting region if it is not 
located at disk center on the stellar surface.  This depends on the exact
topology of the accretion and the limb darkening that should be applied.
The two estimates of $f_{acc}$ are extremely well correlated with each other
and using values estimated from equation (11) produces a correlation in
Figure \ref{vbjf}b very similar to that shown.  However, there are still
worries that $f_{acc}$ and $\dot M$ may show an aritifical correlation,
so testing the relationship given in equation (12) is a better test of
our modified version of the Ostriker and Shu (1995) theory.  Figure \ref{best}
shows the two sides of equation (12) plotted against one another, showing
an excellent correlation.  The resulting correlation coefficient is $r = 0.92$
with an associated false alarm probability of $P_f = 1.5 \times 10^{-7}$.
A least squares analysis finds a best fit line (shown in solid) in the log-log 
plot with a slope of $1.17 \pm 0.12$ which is not significantly different
from the predicted slope of 1.0 (shown in the dashed line).
We are unable to repeat this analysis for CG98 since we do not have access
to their models.  We can do a similar exercise for the data showing the
highest correlation for the other magnetospheric accretion theories.
Examining Table \ref{corr}, the data from HEG may show a possible
correlation when used to test equation (2).  On the other hand, the HEG
data shows a correlation between $\dot M$ and $R_*$ with $r = 0.75$ and
$P_f = 3.1 \times 10^{-4}$.  Again, using equation (10) to eliminate $\dot
M$ from equation (2) gives (after dropping constants)
\begin{equation}
R_*^{97/40} \propto M_*^{11/120} L_{acc}^{23/40} P_{rot}^{29/24}.
\end{equation}
HEG tabulate $L_{acc}$ for each star which we also give in Table
\ref{hegt}.  Using the HEG data to test equation (13) shows only a very
weak correlation, with $r = 0.51$ and $P_f = 0.031$.  The data are shown
in Figure \ref{bestheg} and the slope is $0.88 \pm 0.36$.

In summary, it appears both VBJ and CG98 are consistent with our
extended version of the Ostriker 
and Shu (1995) theory, but the correlation given by CG98 may be the
result of the generally expected correlation between $\dot M$ and $f_{acc} 
R_*^2$.  The data from VBJ actually show the best correlation when considering
all the parameters required by the theory, and the data from this study shows
an excellent correlation with the theory when one tries to remove the
dependence of various parameters on each other.  There is only weak 
support for the more general magnetospheric accretion theories of 
K\"onigl (1991), Cameron and Campbell (1993), Shu et al. (1994), and
Ostriker and Shu (1995) without the extensions described above for 
considering non-dipole field geometeries.

\subsubsection{Filling Factor Versus Stellar Radius}

The above results suggest the Ostriker and Shu (1995) formulation can
be shown to be consistent with the data if the strict dipole field geometry
is dropped and it is assumed that the stellar magnetic field is basically
constant from one CTTS to another.  Ostriker and Shu (1995) state that one 
prediction of their work is that smaller CTTSs should have smaller filling
factors of accretion spots on their surface.  Thus we might expect $f_{acc}$ and
$R_*$ to be directly related to one another.  On the other hand, the
relationship given in equation (7) above clearly predicts that $f_{acc}$ and
$R_*$ should be inversely related to one another.  This apparent contradiction
deserves some comment.

This result can be understood in terms of our equation (6), which was derived
assuming a dipole field geometry in the vicinity of the truncation point of
the disk, but which maps back to an arbitrarily complex field topology at the
stellar surface.  This equation certainly holds for a pure dipole as used by
Ostriker and Shu (1995).  Remembering that $A_* = 4 \pi R_*^2 f_{acc}$, we can
re-write equation (6) as
\begin{equation}
B_* R_*^2 f_{acc} = C_1 M_*^{1/2} \dot M^{1/2} P_{rot}^{1/2}
\end{equation}
where $C_1$ is a constant, $C_1 = 2^{-5/2} \pi^{-1/2} G^{1/2} \Phi_{dx}$.
If we now assume the field is a pure dipole as Ostriker and Shu (1995) do in
their default model, then
\begin{equation}
B_* = \mu_*/R_*^{3+\epsilon}
\end{equation}
where $\mu_*$ is the dipole moment of the star and $\epsilon$ is a small 
positive number ($0 < \epsilon < 1$) which takes into account the fact that
as the stellar radius decreases, the relevant field lines which thread the
disk truncation region in the Ostriker and Shu (1995) model are 
closer to the pole where the field strength is stronger.  
Putting this into equation (14) above yields:
\begin{equation}
\mu_* f_{acc} = R_*^{1+\epsilon} C_1 M*^{1/2} \dot M^{1/2} P_{rot}^{1/2}
\end{equation}
after some re-arrangement.
Holding other things constant, you expect a direct relationship between
$f_{acc}$ and $R_*$ as stated by Ostriker and Shu (1995).  However, in our
extension of the Ostriker and Shu (1995) theory to non-dipole field
geometries at the stellar surface, equation (7) above clearly predicts
an inverse relationship between $f_{acc}$ and $R_*$ if the magnetic field
strength at the stellar surface is held constant along with the stellar
mass, rotation period, and the mass accretion rate.  In a log-log plot
of $f_{acc}$ versus $R_*$, equation (7) predicts a linear relationship with a
slope of -2.

Figure \ref{frstar} shows the relationship between the accretion filling
factor, $f_{acc}$, and the stellar radius, $R_*$, for three data samples.
The top panel uses the data given in Table \ref{cg98t} from CG98.
This panel shows no correlation bewteen the two quantities.  This may simply
reflect the fact that the other relevant parameters ($M_*$, $\dot M$, 
$P_{rot}$) must also be considered simultaneously; however, as mentioned
above, the CG98 analysis does not strictly fit the observations, and in some
case does not match the continuum slopes well.  The middle panel of Figure
\ref{frstar} shows the VBJ data from Table \ref{vbjt}.  In this case an 
inverse correlation is observed: the correlation coefficient is $r = -0.72$
with an associated false alarm probability of $P_f = 1.1 \times 10^{-3}$.  In
a least squares analysis where we again use the distribution around the best
fit line to estimate the uncertainty, the best fit slope is $-1.70 \pm 0.41$,
which is not significantly different from the predicted value of $-2$.
The best fit is shown in a solid line, and the best fit with an assumed
slope of $-2$ is shown in the dashed line.  

One can test this relation using significantly more data points by using
the entire sample of VBJ.  The required filling factors are taken 
from Table 5 of VBJ, and the stellar radii are calculated from the
stellar luminosities given in Table 5 of VBJ and the spectral types given
in Table 1 of VBJ.  Again, to translate spectral type into effective
temperature, we use the calibration given by Johnson (1966).  This entire
data set consists of 43 stars which are plotted in the bottom panel of
Figure \ref{frstar}.  This figure again shows an obvious correlation
which is borne out statistically ($r = 0.69$; $P_f = 9.3 \times 10^{-7}$).
The best fit line (solid) has a slope of $-1.65 \pm 0.28$, which again in
not significantly different from the predicted value of $-2$ (shown in the
dashed line).  We also see that in the entire data sample there is a large
amount of scatter, particularly at the radius corresponding to the stars
in the CG98 sample.  Some of the scatter is likely due to the inherent
variability of CTTS and some may also be due to the fact Figure \ref{frstar}
ignores the effects of other stellar parameters on the correlation.  
Figure \ref{frstar} also illustrates the need for a large data sample when 
making comparisons of this sort.  The fact that we do see the predicted 
correlation in the full data set of VBJ with a slope very close to the 
predicted value suggests that our extension of the Ostriker and Shu (1995) 
theory may be a realistic way to relate the magnetic field in the X-region 
to the stellar surface.  On the other hand, we note that there is an 
observational bias that also predicts $f_{acc} \propto R_*^{-2}$, which is 
shown in equation (11).  For a larger star, a smaller filling factor of 
accretion shock emission is required to give the same area of emitting 
material, and hence the same luminosity, independent of exactly how the 
accreting material falls to the stellar surface.  While the observed 
correlation is completely consistent with the predicition of our extended 
Ostriker and Shu (1995) model, it also suffers from a potential observational
bias.  Figure 6, based on equation (12), remains the strictest test of our 
modification to the Ostriker and Shu (1995) model, and it shows the
best correlation found in this study.

\section{Discussion}

\subsection{Results of This Study}

The model of magnetically controlled accretion onto a compact object is
encountered in many astrophysical contexts.  In the case of CTTSs, such
magnetospheric accretion appears to explain many of the properties of
these stars; however, the general relationships between stellar and
accretion parameters predicted by the theory has not been demonstrated
observationally.
In this paper, we use compilations of rotation periods and studies
of accretion onto CTTSs from 5 sources (VBJ, HEC, GHBC, CG98, WG01) to
test the general predictions of 4 analytic magnetospheric accretion 
models (K\"onigl 1991, Cameron \& Campbell 1993, Shu et al. 1994, and
Ostriker \& Shu 1995).  All the theoretical studies assume a dipole
configuration for the stellar magnetic field; however, Ostriker and
Shu's (1995) detailed analysis of the Shu et al. (1994) theory does not
demand it.  This treatment identifies the essential quantity of the
trapped magnetic flux in the X-region, and this same amount of magnetic 
flux must thread the stellar surface in the accretion zones, independent
of the exact magnetic topology on the stellar surface.

While magnetic field measurements for CTTSs are not numerous, those that
are available show a surprising uniformity of the mean stellar field
strength from star to star (Guenther et al. 1999, Johns--Krull et al. 1999b,
Johns--Krull et al. 2001).  Therefore, we have assumed a constant magnetic
field from star to star and used data from the literature to look
for the predicted correlations among stellar and accretion parameters which
follow from the above mentioned theories.  We suggest that a physical 
reason for the constancy of magnetic fields from one star to the other 
lies in a limit to the efficiency of the dynamo operating in these fully
convective stars.  Generally, the predicted correlations are absent or very
weak in the data.  This is also true of the strict Ostriker and Shu (1995)
theory as applied to a pure dipole magnetic field geometry.  On the other 
hand, Ostriker and Shu (1995) introduce  the concept of the trapped flux, 
which allows one to abandon a strict dipole field geometry and still use 
the theory to make testable predictions (equations 6 and 7).  In order to 
test these predictions, the filling factor, $f_{acc}$, of the accreting 
material on the stellar surface is required.  
 
The studies of VBJ
and CG98 are the only two considered here which present values for $f_{acc}$.
For the stars in common in Tables \ref{vbjt} and \ref{cg98t}, there is
a large range in derived value of $f_{acc}$ between the two studies for some
stars.  However, it is well known that CTTSs are quite variable, showing
both rotational modulation of accretion signatures (e.g. Vrba et al. 1993,
Johns \& Basri 1995, Bouvier et al. 1999) as well as longer timescale
variations in their accretion properties.  In fact, Ardila and Basri (2000)
find variations in the filling factor on BP Tau with a full range from
$f_{acc} = 0.00062 - 0.026$, a factor (42) comparable to the extremes found 
between VBJ and CG98.  This suggests that the filling factor on CTTSs can vary
substantially, which may account for much of the difference in the filling
factors found by VBJ and CG98.  On the other hand, for the stars in Tables 
\ref{vbjt} and \ref{cg98t}, the filling factors derived by VBJ are larger
by a median factor of 3.5 than those found by CG98, which may reflect a 
small systematic difference in the two studies.  For their default model,
Ostriker and Shu (1995) predict a filling factor $f_{acc} = 0.06$, comparable
in value to many of the observed values.  More 
reliable values of the filling factor would provide more stringent tests
of the Ostriker and Shu (1995) model.  G\'omez de Castro and Lamzin (1999)
show that analysis of ultraviolet (UV) emission lines in conjunction with
shock models may result in tight constraints on accretion parameters.
These results are promising, since the UV emission lines arise from the
shock itself, as opposed to the photons being reprocessed through the
photosphere as in the case of the optical excess studied by all five
groups whose results are used in this paper.  Unfortunately, the UV
analysis has not yet proceeded to the point required to test the current
magnetospheric accretion theories.  The data in hand (Figures 1b and 4b) 
show much stronger correlations relative to the pure dipole accretion
theories, with the VBJ data fully consistent with the predictions of
equation (7).  Uncertainties and variations in $f_{acc}$ (as well as the
other relevant parameters including $P_{rot}$) likely contribute to much of
the observed scatter in the plots.

A closer look at the VBJ and CG98 data show that the VBJ data does support
the Ostriker and Shu (1995) model when as many of the interdependencies of 
the various parameters as possible are removed.  The CG98 data suggests that
its apparent support for the extended Ostriker and Shu (1995) model may be 
the result of the expected correlation between the mass accretion rate and 
the area of the star participating in the accretion flow.  On the other hand,
none of the other magnetospheric accretion theories are well supported by the
existing data, making the our extension of the Ostriker and Shu
(1995) description the most viable of the current analytic theories.
We note though that this study by no means rules out the work of K\"onigl
(1991), Cameron and Campbell (1993), or Shu et al. (1994) which employ
strictly dipole field geometries.  It may be that these models are 
essentially correct, and it is the dipole component of the field that
varies from star to star in the manner required by the theory.  We do
note, however, that for the case of BP Tau (Johns--Krull et al. 1999a)
and TW Hya (Johns--Krull \& Valenti 2001) that the {\it measured} dipole
components on these stars are substantially below the theoretically
predicted values.  

\subsection{Implications}

The apparent success of the Ostriker and Shu (1995) trapped flux model once
a dipole geometry is abandoned, and the lack of success for the other 
theories which explicitly assume a dipole field geometry, suggests the 
magnetic fields on CTTSs are not large scale dipoles.  Johns--Krull et al.
(1999b) tabulate dipole field predictions for K\"onigl (1991), Cameron and 
Campbell (1993), and Shu et al. (1994).  Measured field {\it strengths} 
(Basri, Mary, \& Valenti 1992; Guenther et al. 1999, Johns--Krull et al. 
1999b, Johns--Krull et al. 2001) on TTSs are in general agreement with these
predictions; however, the dipole component of these fields appears to be 
smaller than the predictions by a factor of at least 10 in the case of 
K\"onigl (1991) and Shu et al. (1994) based on the lack of detection of 
circular polarization in magnetically sensitive lines of CTTSs (Johns--Krull
1999a, Johns--Krull \& Valenti 2001).  The predicted dipole fields of Cameron
and Campbell (1993) are weaker, but are still ruled out by the observations
at the $6\sigma$ level, and the correlation analysis here does not 
generally support the Cameron and Campbell (1993) formulation.  Independent 
of the theoretical preditions, the measured {\it strength} of the mean 
photospheric fields and the lack of observed circular polarization in the 
magnetically sensitive photospheric lines (Johns--Krull et al. 1999b, 
Johns--Krull \& Valenti 2001) indicates that the surface field topology on 
at least two CTTSs (BP Tau and TW Hya) is not dominted by a dipole.  On the
other hand, the dipole component of the stellar field will fall off the 
slowest with distance from the stellar surface and is likely to dominate at 
the disk truncation radius.  We again note that this generally happens in 
the case of the solar magnetic field at $\sim 2.5 R_\odot$ (Luhmann et al.
1998).  This suggests that the material accreting onto the star generally 
does follow dipole-like magnetic field lines which explains the general 
success of line profile calculations done for dipole geometries (Hartmann, 
Hewett, \& Calvet 1994; Muzerolle et al. 1998, 2001) as well as the circular 
polarization observed in the \ion{He}{1} 5876 \AA\ emission line of several
CTTSs (Johns--Krull et al. 1999a, Johns--Krull \& Valenti 2001).

It is interesting to briefly consider the study of Ferreira et al. (2000)
which explores CTTS interactions with the surrounding disk for both dipole
and higher order magnetic field components.  This study is primarily
concerned with the rotational spin down of young protostars into and through
the TTS phase; however, the authors explicity {\it ignore} the torque 
associated with the surrounding accretion disk.  As a result, the results may
not be accurate.  In all their numerical results, Ferreira et al. (2000) find
a (sometimes steep) increase in the stellar field strength as the star ages
from $10^6$ to $10^7$ years and beyond.  The limited magnetic field 
observations available do not currently support this picture: Johns--Krull
and Valenti (2001) find a mean magnetic field on TW Hya (age $10^7$ yr -
Webb et al. 1999) which is well within the range found for CTTSs is 
Taurus.  Additionally, Valenti and Johns--Krull (2001) report a mean field
on a K3V Pleiades (age $\sim 10^8$ yr - Basri, Marcy, \& Graham 1996) member
which is {\it less} than those of TTS by about a factor of 2.  More data is 
needed, particularly on young stars with an age of $\sim 10^7$ yr to see if 
any trends emerge.

We have just argued that the magnetic fields of CTTSs are not large scale
dipoles and that the dipole component of CTTSs is significantly smaller
than required by the magnetospheric accretion models of K\"onigl (1991),
Cameron \& Campbell (1993), and Shu et al. (1994).  Ostriker and Shu (1995)
detail the theory of Shu et al. (1994) and show that the dipole assumtion
is not necessary by introducing the idea of trapped flux.  How does the
field strength in the trapped flux model compare with the observed field
strength?  Solving equation (6) for the stellar magnetic field gives:
\begin{equation}
B_* = 14,500 \Bigl({M_* \over M_\odot}\Bigr)^{1/2}
\Bigl({\dot M_D \over 10^{-7} M_\odot\,\, {\rm yr^{-1}}}\Bigr)^{1/2}
\Bigl({P_{rot} \over 1\, {\rm day}}\Bigr)^{1/2} 
\Bigl({R_* \over R_\odot}\Bigr)^{-2} 
\Bigl({f_{acc} \over 0.01}\Bigr)^{-1} \,\,\, {\rm G}
\end{equation}
where we have set $A_* = 4 \pi R_*^2 f_{acc}$.  Since only the studies of VBJ
and CG98 independently estimate the filling factor, $f_{acc}$, required in this 
equation, we only use these studies to estimate the stellar magnetic field
strength, $B_*$.  Table \ref{magp} gives these estimates for the 3 stars
observed by Johns--Krull et al. (2001) for which the calculation can be made.
Also given in the table is the observed mean field strength from Johns--Krull
et al. (2001).  It is important to note that in the analysis of Johns--Krull
et al. (1999b, 2001) evidence is found for a range of field strengths
on the surface of CTTSs, and we are unable to specify the field strength
only in the accretion zone.  The \ion{He}{1} polarization data of Johns--Krull
and Valenti (2000) and Valenti et al. (2001) probably probes the field in
the accretion zone at the stellar surface, but there is an unknown inclination
of this field to the line of sight which reduces the observed field strength
below the actual field value.  With these caveats, the predicted and
observed field strengths agree to within a factor of 2-3, with the predictions
using the parameters of VBJ generally falling below those based on the
parameters of CG98 in this sample of 3 stars.  Except for the rotation
period, these two studies find different values for all the relevant
parameters in equation (11); however, the one which is most significant 
in producing the difference in the predicted field strengths is the filling
factor, $f_{acc}$.  CTTSs are know to
vary due to rotational modulation.  In addition, there are also likely to
be real changes in the instantaneous accretion rate onto the star (e.g.
Ardila \& Basri 2000).  Both processes could change the apparent filling
factor of accreting material.  As a result (and as mentioned above), it is 
not clear that the difference in the derived filling factor values between 
VBJ and CG98 is significant or simply the result of the variability of these 
stars. 

    Finally, we note that the relationships set forth in this paper are based 
on equilibrium calculations, while CTTSs do show substantial variation.  As
a result, we expect that the relationships should hold for the mean
value of the relevant parameters (particularly the accretion rate), but
probably do not describe the relationship between these parameters as they
vary over relatively short timescales (compared to the evolutionary
timescale of CTTSs and their disks).  For example, Ardila and Basri (2000)
study the accretion variability of the CTTSs BP Tau, finding that 
$f_{acc} \propto \dot M^2$.  At first, this appears to contradict equation (7).
Rotational modulation may produce some of the variation observed by 
Ardila and Basri (2000).  Aside from this, over the timescale of the 
variations in the instantaneous accretion rate studied by Ardila and Basri 
(2000), it is doubtful that the star reached a new equilibrium rotation rate 
which it would be expected to do.  In other words, the torque balance on the
star may also be varying, alternately trying to speed it up or slow it down, 
all the time varying around the equilibrium value.  We would only expect
equation (7) to hold for the mean values, as it seems to based on the
correlation plots presented in this paper.  Indeed, some of the scatter
seen in these plots is likely to result from the intrinsic variability
displayed by CTTSs.

\section{Conclusions}

The magnetospheric accretion model can explain many observed features of
CTTSs.  Equilibrium models of this process have relied primarily on dipole
field geometries to explore the interaction of the star with the disk, which
seems appropriate since the dipole component should dominate at large
distance from the star.  Observations suggest that the magnetic fields
of CTTSs do not change much from star to star, though further observations
are needed to confirm this.  Under the assumption that the fields do not
vary strongly, equilibrium theories predict specific correlations between
stellar and accretion parameters which are generally not observed.  This may
mean that equilibrium theories are simply not adequate, or that a dipole
field topology is too simple an assumption for relating observable parameters.
On the other hand, we can use the notion of trapped flux introduced by
Ostriker and Shu (1995) to extend this equilibrium theory to non-dipole field 
topologies at the stellar surface.  The extension we make relies on
assuming that the dipole component of the magnetic field does dominate
at the X-region in the Ostriker and Shu (1995) model and that their analysis
is accurate in terms of the accounting of how much magnetic flux is in the
accretion flow, the dead zone, and the wind.  This determines the magnetic
flux in the accretion flow which we assume maps back to an equal amount of
magnetic flux on the stellar surface.  Again holding the magnetic field
constant, new correlations are predicted bewteen stellar and accretion
parameters.  We argue that the data of VBJ are the best suited for looking
for these correlations, and that they are apparent in the data.  This
suggests that equilibrium models are adequate to describe the general
features of magnetospheric accretion in CTTSs once the dipole field assumption
is dropped.  Additional tests of our modified Ostriker and Shu (1995) model 
will confirm or refute this.  Particularly valuable tests would be to 
determine all the relevant stellar and accretion parameters for stars with 
substantially different stellar properties than the Taurus stars used here.
For example, detailed observations (including magnetic field measurements) of
the more rapidly rotating CTTSs in Orion could provide critical tests of 
equation (6) and (7).  Tests of these equations will also benefit from better
estimates of accretion rates and filling factors which may result from
the analysis of UV emission lines produced in the accretion shock (e.g.
G\'omez de Castro \& Lamzin 1999).  Until then, it appears that our
modified Ostriker and Shu (1995) is a reasonable description of magnetospheric
accretion onto CTTSs.

\acknowledgements
CMJ-K would like to acknowledge stimulating discussions with F. Shu and to
thank him for pointing out the role of the trapped flux.  We 
acknowledge the careful reading of the manuscript by S. Mohanty and the
useful discussions which resulted.  Communications with L. Prato and R.
White also provided important input on several issues raised in the 
manuscript.  We also wish to thank an anonymous referee for feedback 
provided on this paper.  CMJ-K and ADG both acknowledge partial support from
a NASA Origins grant NAG5-8098 to the Regents of the University of California.

\clearpage

\figcaption{(a) The top panel shows the quantity $(M_*/M_\odot)^{5/6}
(\dot M/1 \times 10^{-7} M_\odot yr^{-1})^{1/2} P_{rot}^{7/6}$
versus $(R_*/R_\odot)^3$ for the sample of stars from VBJ.  Single CTTSs
are shown in solid circels while CTTSs in binary systems are shown in
asterisks.  The dashed line shows the best fit line whose slope (1.0) is
predicted by equation (1).
(b) The bottom panel shows the quantity $(M_*/M_\odot)^{1/2}
(\dot M/1 \times 10^{-7} M_\odot yr^{-1})^{1/2} P_{rot}^{1/2}$
versus $(R_*/R_\odot)^2 f_{acc}$ for the sample of stars from VBJ.  Plot
symbols are the same as in panel (a).  Shown in the 
solid line is the best fit line to the data, and shown in the dashed line
is best fit line whose slope (1.0) is predicted by equation (7).
\label{vbjf}}

\figcaption{Same as Figure 1a but using the data of HEG.  Also shown in
the solid curve is the best fit line to the data.
\label{hegf}}

\figcaption{Same as Figure 1a but using the data of GHBC.  
\label{ghbcf}}

\figcaption{Same as Figure 1 but using the data of CG98.
\label{cg98f}}

\figcaption{Same as Figure 1a but using the data of WG01.
\label{wg01f}}

\figcaption{The data from VBJ are used to test for the correlation
predicted by equation (12).  The dashed line shows the best fit match to
the points keeping the slope fixed to the value predicted in equation
(12), while the best fit line with a free slope is shown in the solid
line.
\label{best}}

\figcaption{The data from HEG are used to test for the correlation
predicted by equation (13).  The dashed line shows the best fit match to
the points keeping the slope fixed to the value predicted in equation
(13), while the best fit line with a free slope is shown in the solid
line.
\label{bestheg}}

\figcaption{The accretion filling factor, $f_{acc}$, versus the stellar radius,
$R_*$, for three data samples.  The upper panel uses the data of CG98 as given
in this paper.  The middle panel is data from VBJ as given in this paper, and
the lower panel is the full set of VBJ data.  Again, single CTTSs are shown as
solid circles and those in binary systems are shown as asterisks.  In each
panel, the dashed line is the best fit line with an assumed slope of
-2 (the predicted slope).  In the lower two panels where real correlations
do exist, the best fit line is plotted in the solid line.  In these two
panels, the best fit line is not significantly different from the expected
relationship.
\label{frstar}}

\newpage
\epsscale{.75}
\plotone{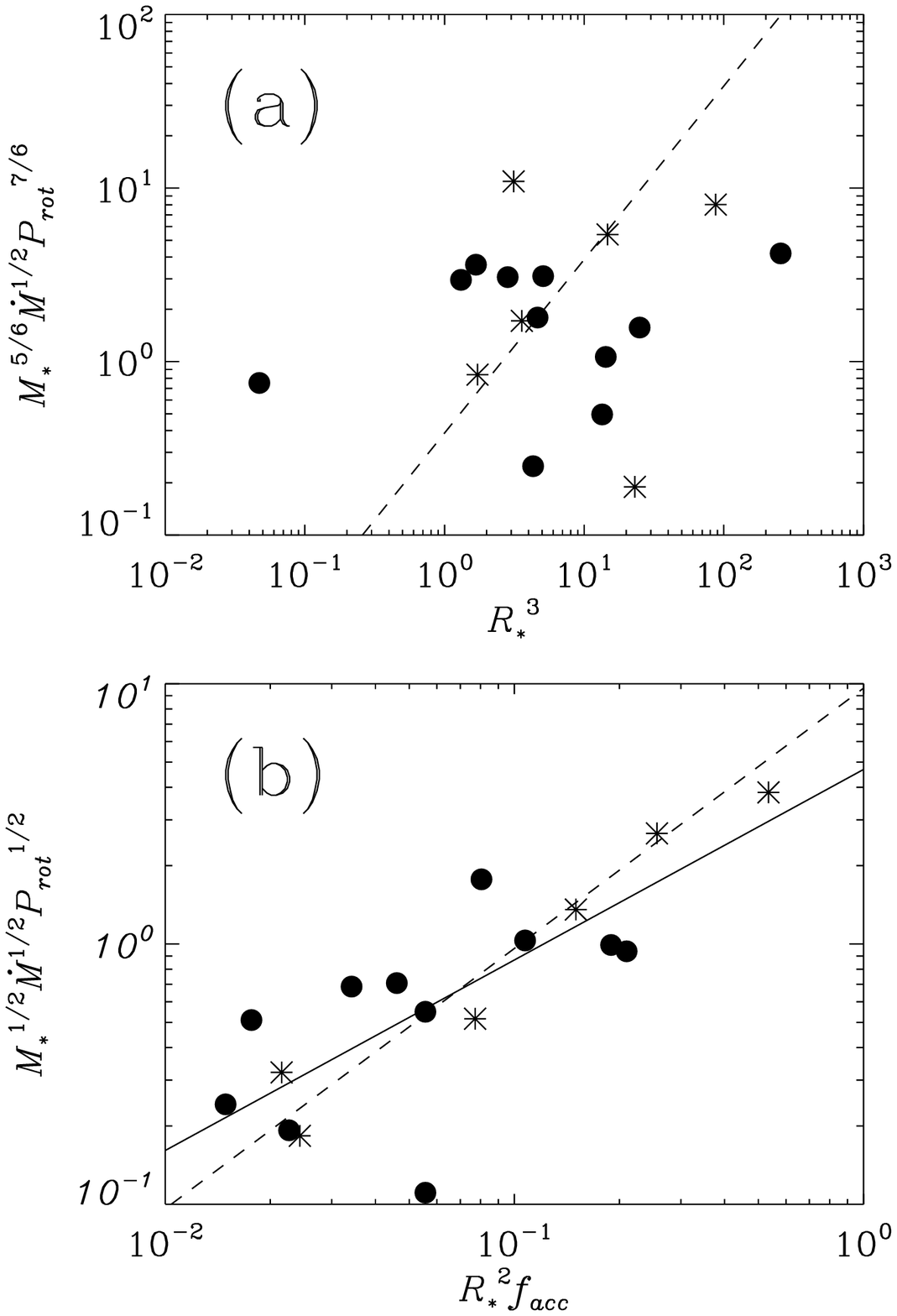}
\newpage
\plotone{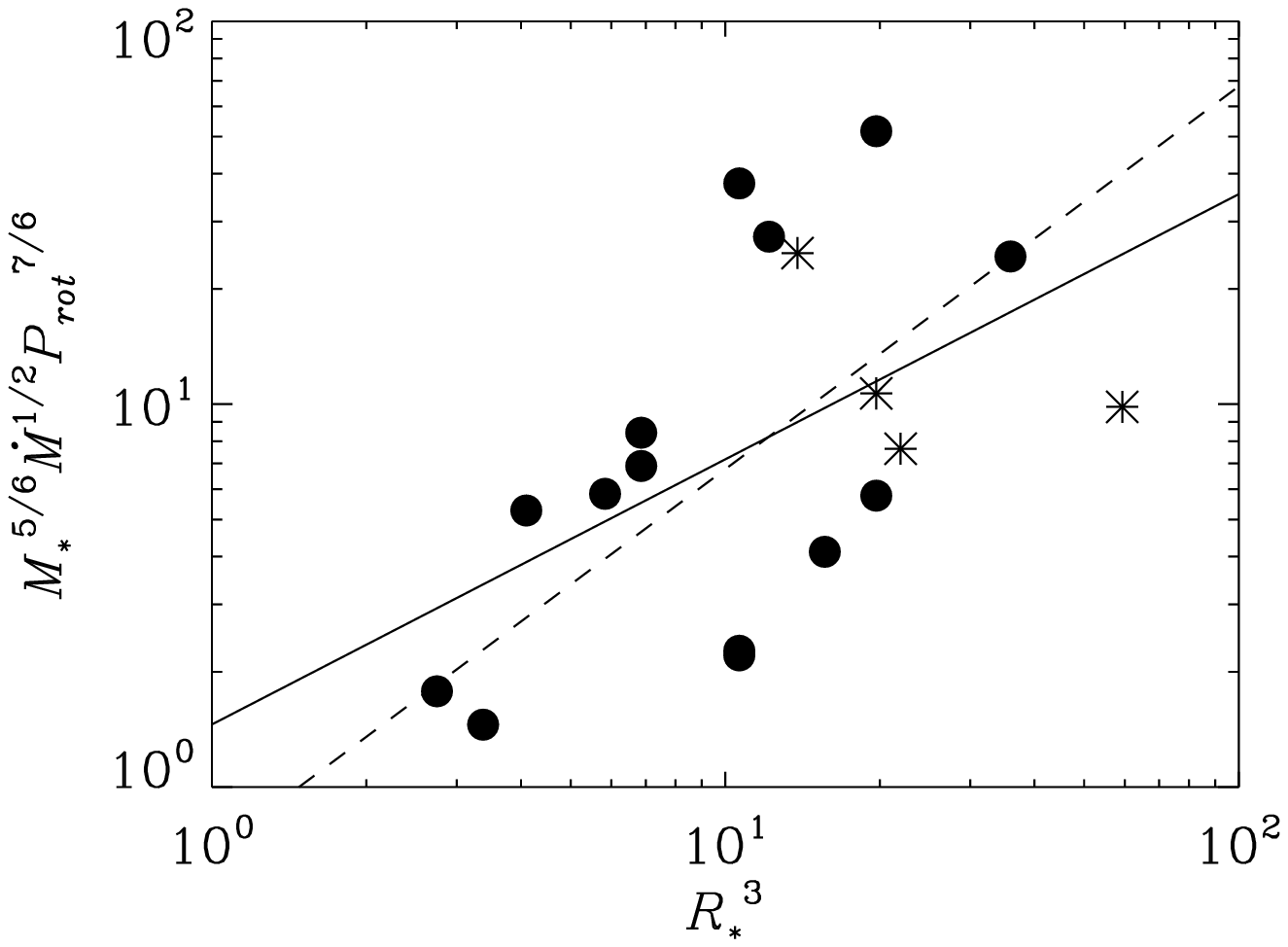}
\newpage
\plotone{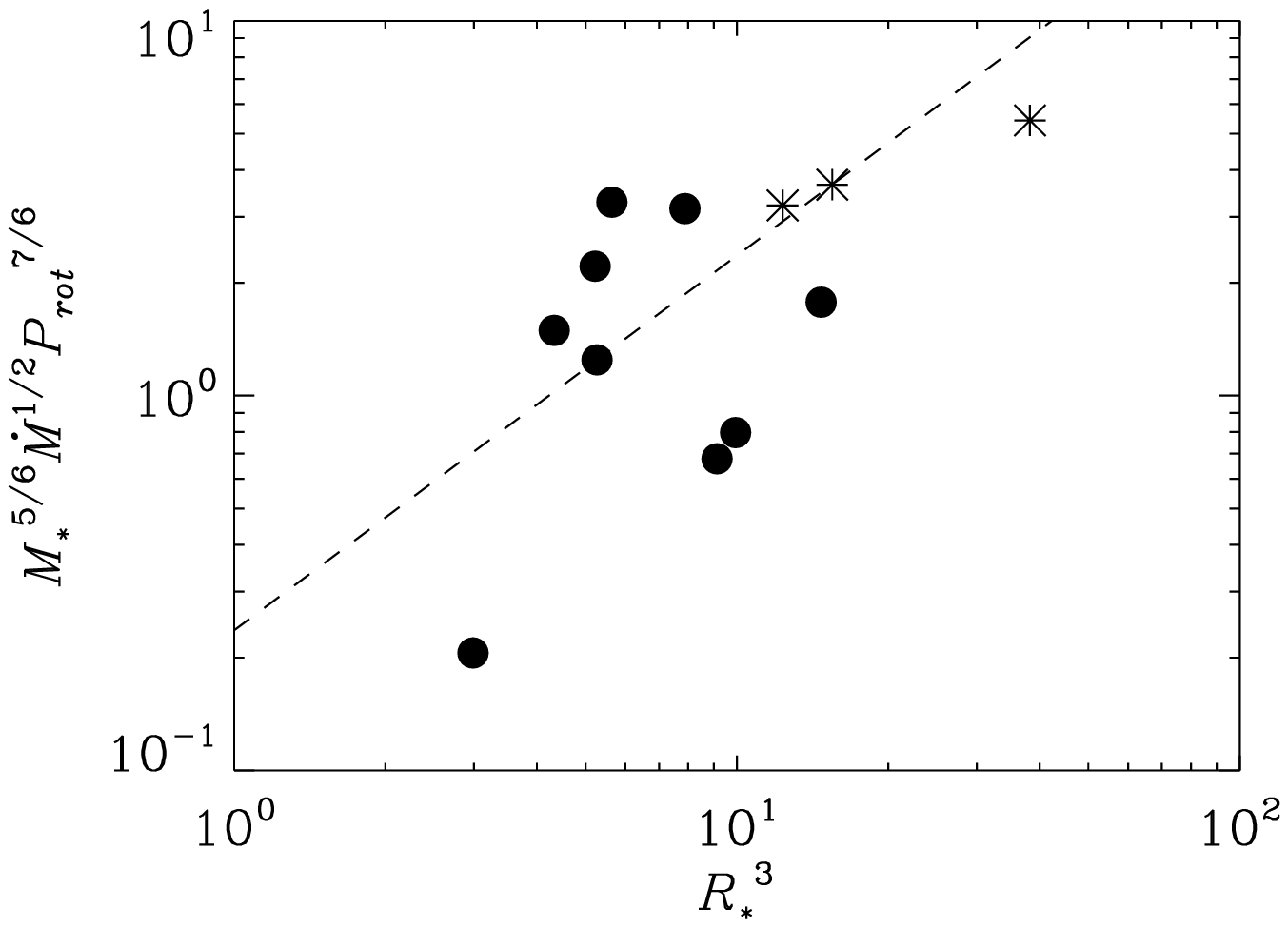}
\newpage
\plotone{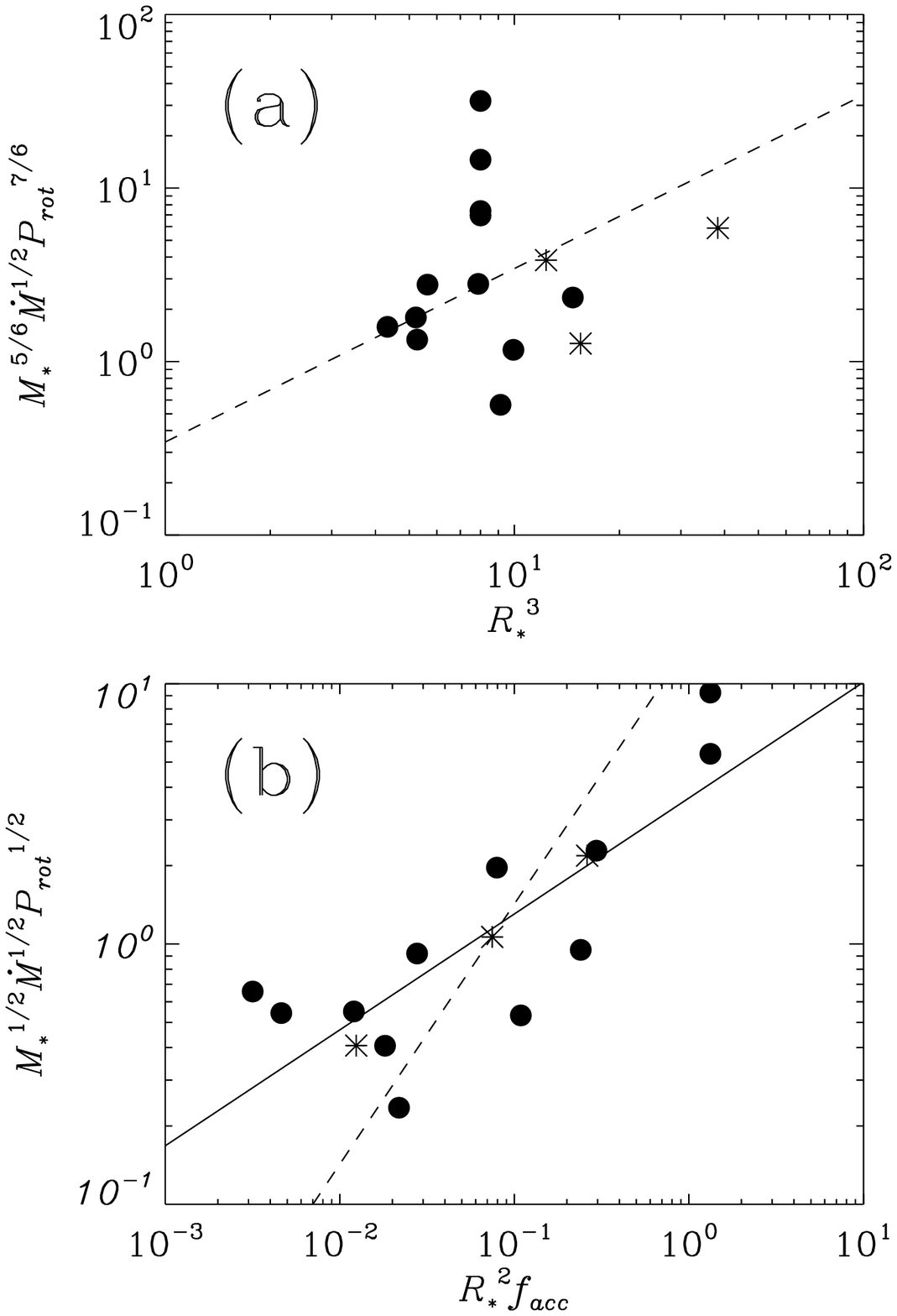}
\newpage
\plotone{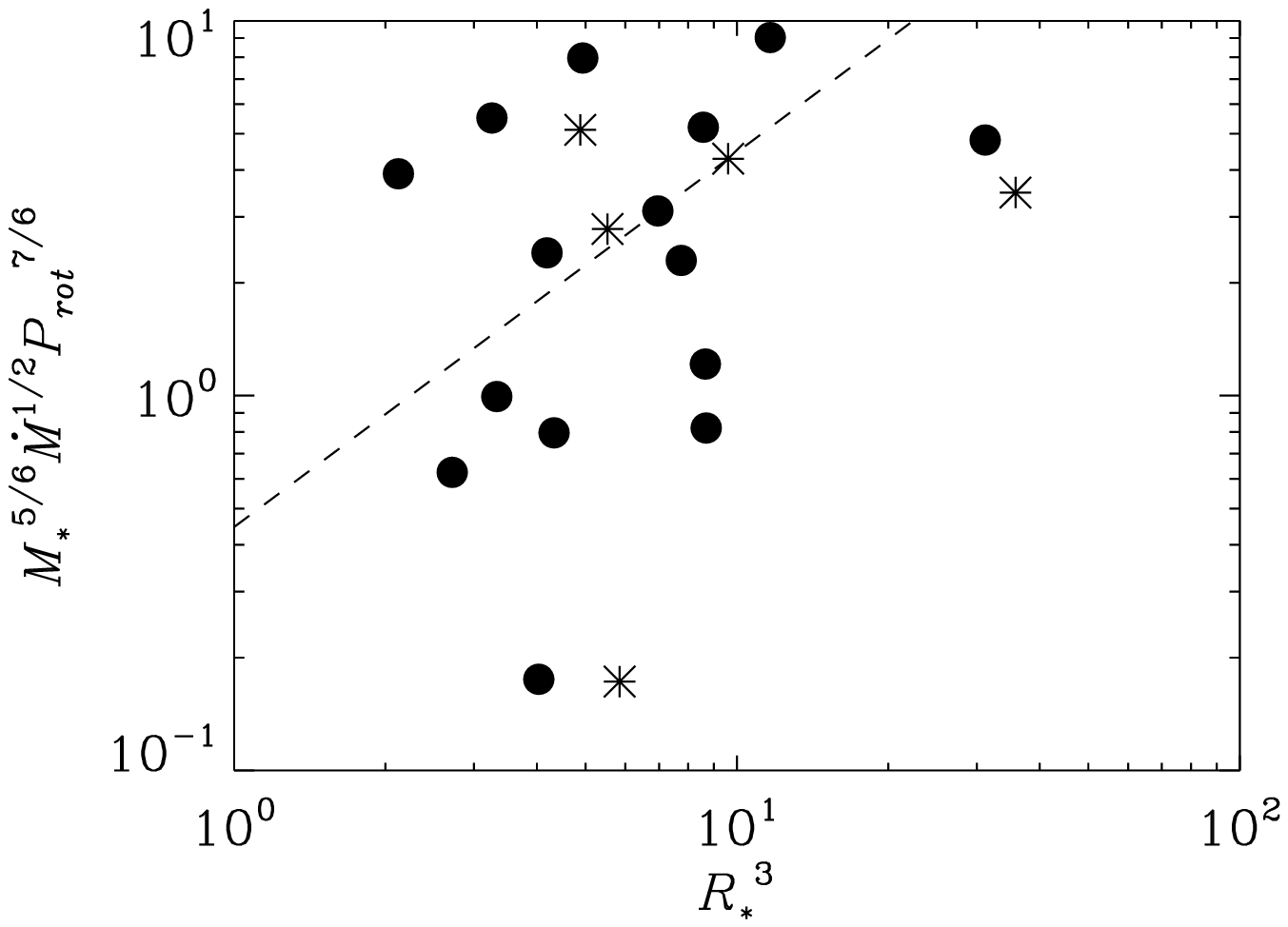}
\newpage
\plotone{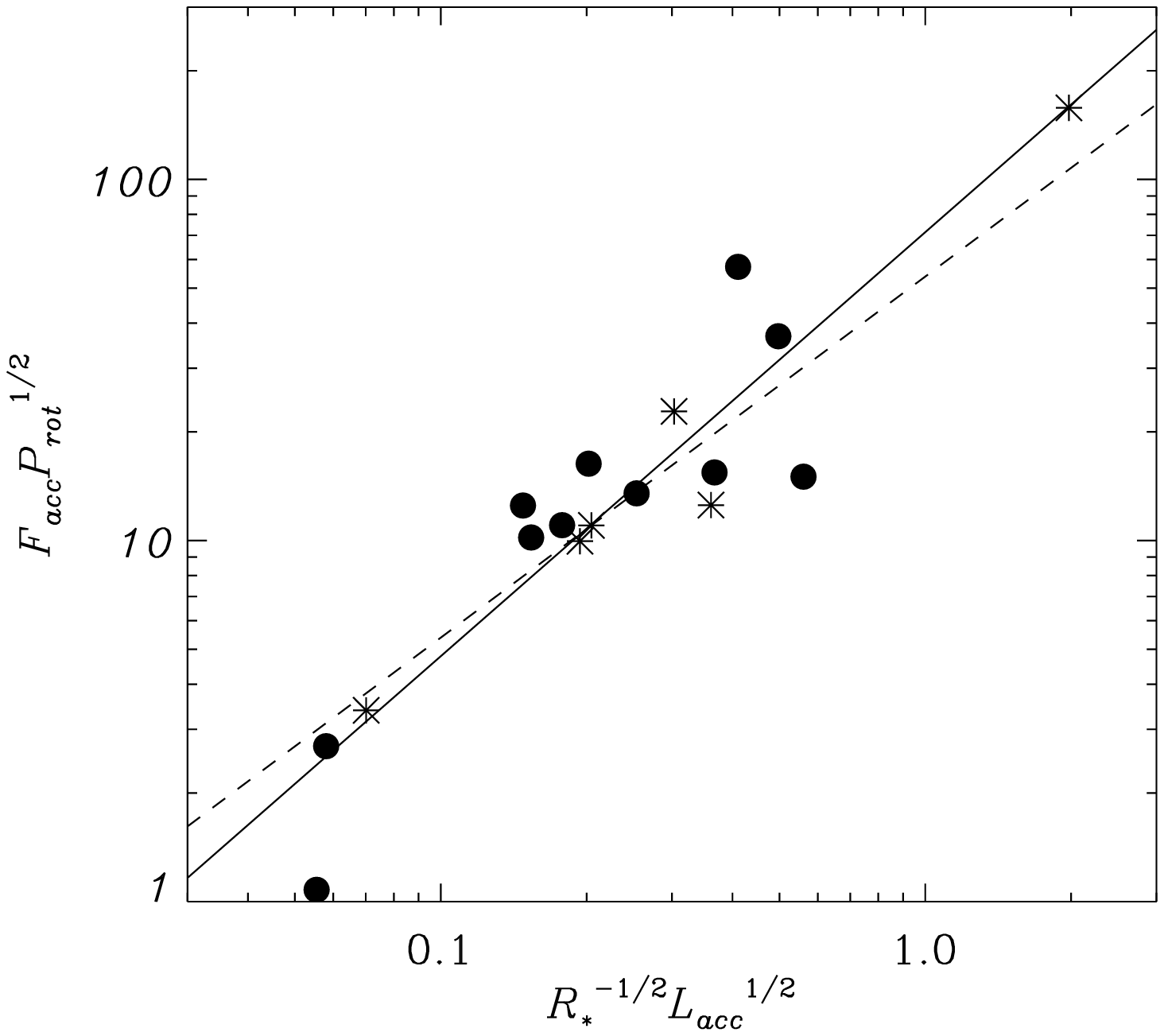}
\newpage
\plotone{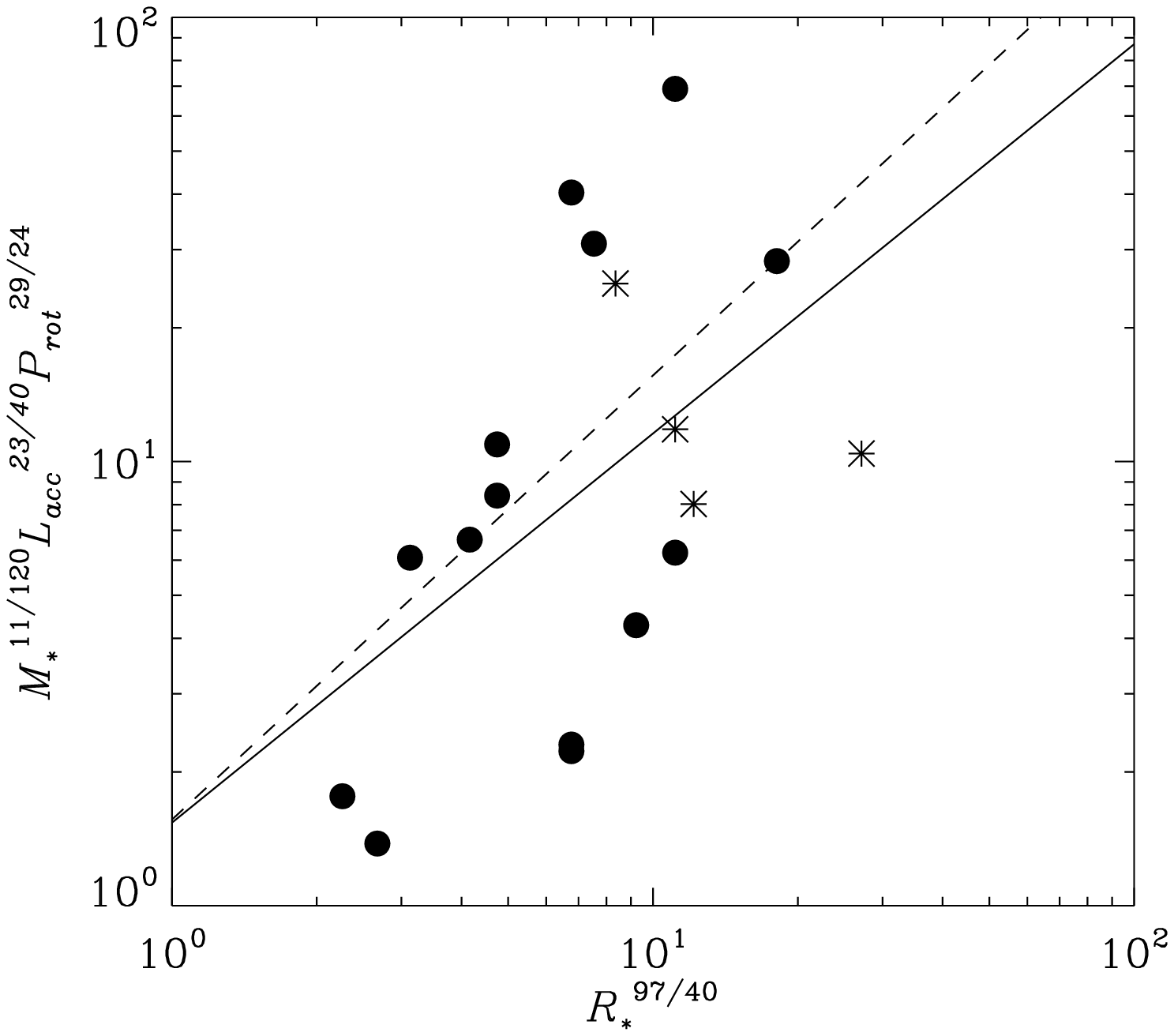}
\newpage
\plotone{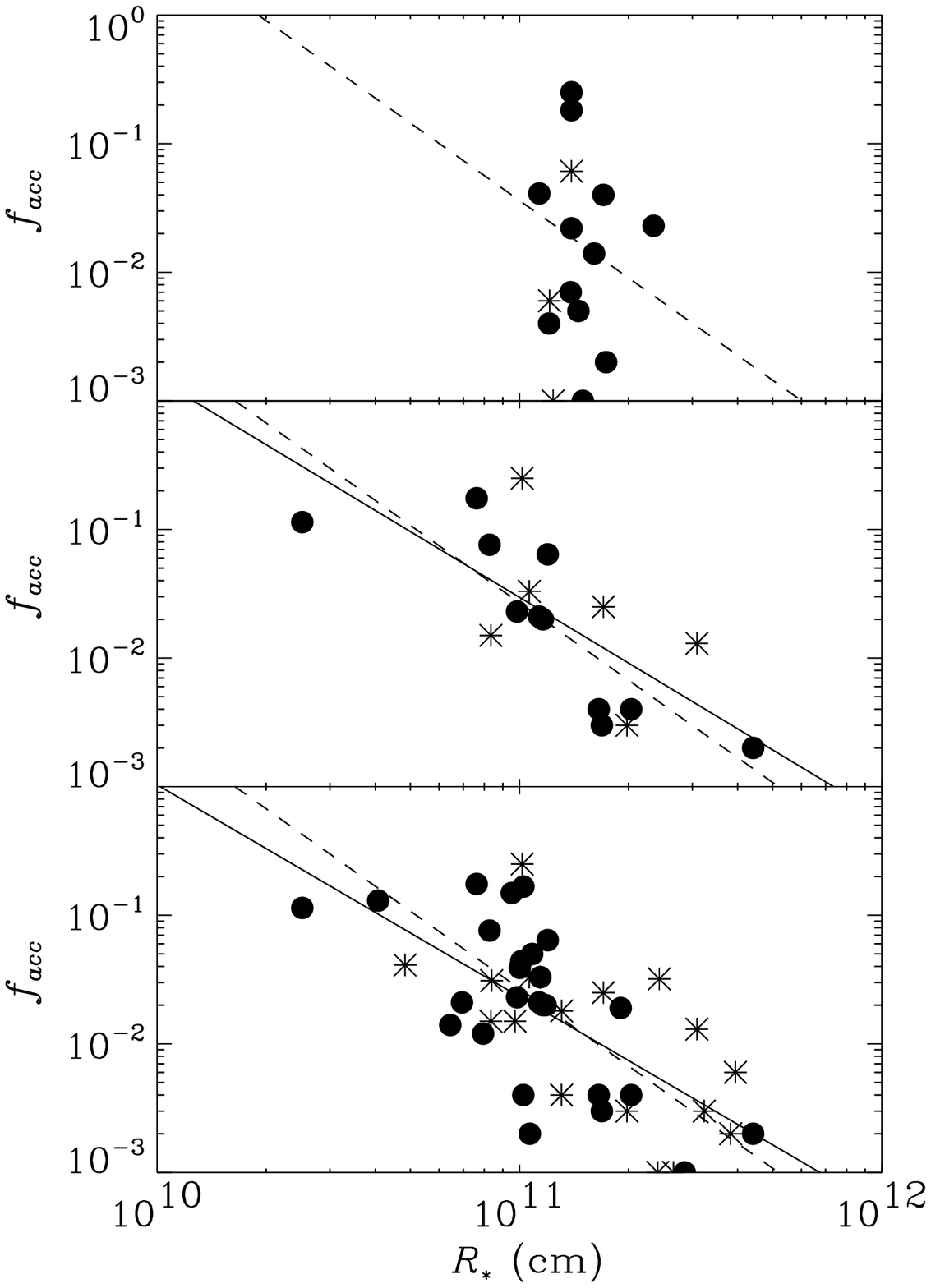}

\clearpage
 
\begin{deluxetable}{lcccccccc}
\tablewidth{16.0truecm}   
\tablecaption{VBJ Sample of Stars\label{vbjt}}
\tablehead{
   \colhead{}&
   \colhead{$T_{eff}$}&
   \colhead{$M_*$}&
   \colhead{$R_*$\tablenotemark{a}}&
   \colhead{$\dot{M} \times 10^7$}&
   \colhead{}&
   \colhead{$P_{rot}$}&
   \colhead{}&
   \colhead{$F_{acc} \times 10^{-10}$}\\[0.2ex]
   \colhead{Star}&
   \colhead{(K)}&
   \colhead{(M$_\odot$)}&
   \colhead{(R$_\odot$)}&
   \colhead{$(M_\odot {\rm yr}^{-1})$}&
   \colhead{$f_{acc}$}&
   \colhead{(days)}&
   \colhead{${L_{acc} \over L_\odot}$}&
   \colhead{(erg cm$^{-2}$ s$^{-1}$)}
}
\startdata
AA Tau & 4000 &   0.52 & 1.67 & 0.071 & 0.020 & 8.20 & 0.068 & 5.7 \\
BP Tau & 4000 &   0.53 & 1.72 & 0.243 & 0.064 & 7.60 & 0.232 & 5.6 \\
CW Tau & 4775 &   0.45 & 0.36 & 0.016 & 0.114 & 8.20 & 0.061 & 20. \\
DE Tau & 3400 &   0.23 & 6.34 & 1.796 & 0.002 & 7.60 & 0.201 & 4.0 \\
DF Tau\tablenotemark{b} & 3675 &  0.38 & 4.43 & 2.192 & 0.013 & 8.50 & 
  0.579 & 4.3 \\
DH Tau & 3600 &   0.23 & 2.92 & 0.283 & 0.004 & 7.20 & 0.069 & 3.8 \\
DK Tau\tablenotemark{b} & 4000 &   0.52 & 1.53 & 0.061 & 0.033 & 8.40 & 
  0.064 & 3.8 \\
DL Tau & 4000 &   0.49 & 1.19 & 0.233 & 0.076 & 9.40 & 0.294 & 12. \\
DN Tau & 3750 &   0.47 & 2.39 & 0.013 & 0.004 & 6.00 & 0.008 & 1.1 \\
GG Tau A\tablenotemark{b} & 4000 &   0.59 & 2.45 & 0.301 & 0.025 & 10.30 & 
  0.225 & 7.1 \\
GI Tau & 4200 &   0.60 & 1.09 & 0.202 & 0.175 & 7.20 & 0.344 & 5.6 \\
GK Tau & 4000 &   0.53 & 1.63 & 0.005 & 0.021 & 4.65 & 0.005 & 0.5 \\
GM Aur & 4775 &   0.56 & 1.42 & 0.074 & 0.023 & 12.00 & 0.091 & 3.9 \\
HP Tau\tablenotemark{b} & 4775 &   0.51 & 1.20 & 0.034 & 0.015 & 5.90 & 
  0.045 & 4.1 \\
IQ Tau & 3675 & 0.23 & 2.43 & 0.180 & 0.003 & 6.25 & 0.053 & 5.0 \\
RW Aur AB\tablenotemark{b} & 5100 &  0.80 & 1.46 & 3.384 & 0.250 & 5.39 & 
  5.725 & 68. \\
XZ Tau\tablenotemark{b} & 3300 &   0.16 & 2.85 & 0.079 & 0.003 & 2.60 & 
  0.014 & 2.1 \\
\enddata
\tablenotetext{a}{Calculated from stellar luminosity and effective
temperature.}
\tablenotetext{b}{Denotes a star in a binary system.}
\end{deluxetable}

\clearpage

\begin{deluxetable}{lcccccc}
\tablewidth{12.0truecm}   
\tablecaption{HEG Sample of Stars\label{hegt}}
\tablehead{
   \colhead{}&
   \colhead{$T_{eff}$}&
   \colhead{$M_*$}&
   \colhead{$R_*$}&
   \colhead{$\dot{M} \times 10^7$}&
   \colhead{$P_{rot}$}&
   \colhead{}\\[0.2ex]
   \colhead{Star}&
   \colhead{(K)}&
   \colhead{(M$_\odot$)}&
   \colhead{(R$_\odot$)}&
   \colhead{$(M_\odot {\rm yr}^{-1})$}&
   \colhead{(days)}&
   \colhead{log$\bigl({L_{acc} \over L_\odot}\bigr)$}
}
\startdata
AA Tau & 3800 & 0.38 & 1.8 & 1.26 & 8.20 & $-$0.42 \\
BP Tau & 4000 & 0.45 & 1.9 & 1.58 & 7.60 & $-$0.19 \\
CW Tau & 4730 & 1.03 & 2.2 & 10.00 & 8.20 & 0.87 \\
CY Tau & 4000 & 0.58 & 1.4 & 0.06 & 7.90 & $-$1.42 \\
DE Tau & 3570 & 0.24 & 2.7 & 3.16 & 7.60 & $-$0.37 \\
DF Tau\tablenotemark{a} & 3500 & 0.17 & 3.9 & 12.59 & 8.50 & $-$0.06 \\
DG Tau & 4395 & 0.67 & 2.3 & 19.95 & 6.30 & 0.94 \\
DK Tau\tablenotemark{a} & 4000 & 0.38 & 2.7 & 3.98 & 8.40 & $-$0.01 \\
DL Tau & 3800 & 0.37 & 1.9 & 2.00 & 9.40 & $-$0.17 \\
DN Tau & 4000 & 0.42 & 2.2 & 0.32 & 6.00 & $-$0.97 \\
DR Tau & 4000 & 0.38 & 2.7 & 79.43 & 9.00 & 1.26 \\
GG Tau A\tablenotemark{a} & 3800 & 0.29 & 2.8 & 2.00 & 10.30 & $-$0.47 \\
GI Tau & 3800 & 0.30 & 2.5 & 1.26 & 7.20 & $-$0.62 \\
GK Tau & 4000 & 0.41 & 2.2 & 0.63 & 4.65 & $-$0.71 \\
GM Aur & 4000 & 0.52 & 1.6 & 0.25 & 12.00 & $-$0.86 \\
RW Aur\tablenotemark{a} & 4000 & 0.85 & 2.4 & 15.85 & 5.39 & 0.91 \\
V836 Tau & 4000 & 0.54 & 1.5 & 0.06 & 7.00 & $-$1.49 \\
YY Ori & 4000 & 0.34 & 3.3 & 31.62 & 7.58 & 0.75 \\
\enddata
\tablenotetext{a}{Denotes a star in a binary system.}
\end{deluxetable}

\clearpage

\begin{deluxetable}{lccccc}
\tablewidth{10.5truecm}   
\tablecaption{GHBC Sample of Stars\label{ghbct}}
\tablehead{
   \colhead{}&
   \colhead{$T_{eff}$}&
   \colhead{$M_*$}&
   \colhead{$R_*$}&
   \colhead{$\dot{M} \times 10^7$}&
   \colhead{$P_{rot}$}\\[0.2ex]
   \colhead{Star}&
   \colhead{(K)}&
   \colhead{(M$_\odot$)}&
   \colhead{(R$_\odot$)}&
   \colhead{$(M_\odot {\rm yr}^{-1})$}&
   \colhead{(days)}
}
\startdata
 AA Tau & 3800 &  0.53 & 1.74 &  0.033 & 8.20 \\
 BP Tau & 4000 &  0.49 & 1.99 &  0.288 & 7.60 \\
 CY Tau & 4000 &  0.42 & 1.63 &  0.075 & 7.90 \\
 DE Tau & 3570 &  0.26 & 2.45 &  0.264 & 7.60 \\
 DF Tau\tablenotemark{a} & 3500 &  0.27 & 3.37 &  1.769 & 8.50 \\
 DK Tau\tablenotemark{a} & 4000 &  0.43 & 2.49 &  0.379 & 8.40 \\
 DN Tau & 4000 &  0.38 & 2.09 &  0.035 & 6.00 \\
 GG Tau A\tablenotemark{a} & 3800 &  0.44 & 2.31 &  0.175 & 10.30 \\
 GI Tau & 3800 &  0.67 & 1.74 &  0.096 & 7.20 \\
 GK Tau & 4000 &  0.46 & 2.15 &  0.064 & 4.65 \\
 GM Aur & 4000 &  0.52 & 1.78 &  0.096 & 12.00 \\
 IP Tau & 3750 &  0.52 & 1.44 &  0.008 & 3.25 \\
\enddata
\tablenotetext{a}{Denotes a star in a binary system.}
\end{deluxetable}

\clearpage

\begin{deluxetable}{lcccccc}
\tablewidth{11.5truecm}   
\tablecaption{CG98 Sample of Stars\label{cg98t}}
\tablehead{
   \colhead{}&
   \colhead{$T_{eff}$\tablenotemark{a}}&
   \colhead{$M_*$\tablenotemark{a}}&
   \colhead{$R_*$\tablenotemark{a}}&
   \colhead{$\dot{M} \times 10^7$\tablenotemark{b}}&
   \colhead{}&
   \colhead{$P_{rot}$}\\[0.2ex]
   \colhead{Star}&
   \colhead{(K)}&
   \colhead{(M$_\odot$)}&
   \colhead{(R$_\odot$)}&
   \colhead{$(M_\odot {\rm yr}^{-1})$}&
   \colhead{$f_{acc}$}&
   \colhead{(days)}
}
\startdata
AA Tau & 3800 & 0.53  & 1.74 & 0.04 & 0.006 & 8.20 \\
BP Tau & 4000 & 0.49  & 1.99 & 0.23 & 0.007 & 7.60 \\
CW Tau & 4730 & 0.50  & 2.00 & 1.27  & 0.061 & 8.20 \\
CY Tau & 4000 & 0.42  & 1.63 & 0.08 & 0.041 & 7.90 \\
DE Tau & 3570 & 0.26  & 2.45 & 0.46 & 0.040 & 7.60 \\
DF Tau\tablenotemark{c} & 3500 & 0.27  & 3.37 & 2.08 & 0.023 & 8.50 \\
DG Tau & 4395 & 0.50  & 2.00 & 9.19  & 0.251 & 6.30 \\
DK Tau\tablenotemark{c} & 4000 & 0.43  & 2.49 & 0.05 & 0.002 & 8.40 \\
DL Tau & 3800 & 0.50  & 2.00 & 0.82  & 0.022 & 9.40 \\
DN Tau & 4000 & 0.38  & 2.09 & 0.02 & 0.005 & 6.00 \\
DR Tau & 4000 & 0.50  & 2.00 & 19.01 & 0.182 & 9.00 \\
GG Tau A\tablenotemark{c} & 3800 & 0.44  & 2.31 & 0.25 & 0.014 & 10.30 \\
GI Tau & 3800 & 0.67  & 1.74 & 0.06 & 0.004 & 7.20 \\
GK Tau & 4000 & 0.46  & 2.15 & 0.14 & 0.001 & 4.65 \\
GM Aur & 4000 & 0.52  & 1.78 & 0.07 & 0.001 & 12.00 \\
\enddata
\tablenotetext{a}{Taken from GHBC except for CW Tau, DL Tau, DG Tau, and
DR Tau which are assumed by CG98 to have $M_* = 0.5 M_\odot$ and $R_* = 
2.0 R_\odot$.}
\tablenotetext{b}{Calculated from equation (11) of CG98.}
\tablenotetext{c}{Denotes a star in a binary system.}
\end{deluxetable}

\clearpage

\begin{deluxetable}{lccccc}
\tablewidth{10.5truecm}   
\tablecaption{WG01 Sample of Stars\label{wg01t}}
\tablehead{
   \colhead{}&
   \colhead{$T_{eff}$}&
   \colhead{$M_*$}&
   \colhead{$R_*$\tablenotemark{a}}&
   \colhead{$\dot{M} \times 10^7$}&
   \colhead{$P_{rot}$}\\[0.2ex]
   \colhead{Star}&
   \colhead{(K)}&
   \colhead{(M$_\odot$)}&
   \colhead{(R$_\odot$)}&
   \colhead{$(M_\odot {\rm yr}^{-1})$}&
   \colhead{(days)}
}
\startdata
AA Tau & 4000 & 0.78 & 1.61 & 0.06 & 8.20 \\
BP Tau & 4000 & 0.77 & 1.91 & 0.13 & 7.60 \\
CW Tau & 4775 & 1.06 & 1.28 & 0.10 & 8.20 \\
CY Tau & 3400 & 0.55 & 1.63 & 0.01 & 7.90 \\
DE Tau & 3600 & 0.66 & 3.15 & 0.41 & 7.60 \\
DF Tau A\tablenotemark{b} & 3775 & 0.68 & 1.77 & 0.10 & 8.50 \\
DG Tau & 4200 & 0.88 & 2.05 & 0.07 & 6.30 \\
DH Tau & 3400 & 0.53 & 1.39 & 0.01 & 7.20 \\
DL Tau & 4000 & 0.77 & 2.27 & 0.68 & 9.40 \\
DN Tau & 3750 & 0.70 & 2.06 & 0.02 & 6.00 \\
DR Tau & 4400 & 1.11 & 1.70 & 0.32 & 9.00 \\
GG Tau Aa\tablenotemark{b} & 4000 & 0.76 & 2.13 & 0.13 & 10.3 \\
GI Tau & 4000 & 0.76 & 1.98 & 0.08 & 7.20 \\
GK Tau & 4000 & 0.76 & 2.05 & 0.06 & 4.65 \\
GM Aur & 4775 & 1.22 & 1.48 & 0.07 & 12.00 \\
IP Tau & 3750 & 0.71 & 1.59 & 0.003 & 3.25 \\
LkCa 15 & 4400 & 1.05 & 1.49 & 0.01 & 5.85 \\
RW Aur A\tablenotemark{b} & 5080 & 1.34 & 1.70 & 0.32 & 5.39 \\
T Tau A\tablenotemark{b} & 5250 & 2.11 & 3.30 & 0.32 & 2.80 \\
XZ Tau A\tablenotemark{b} & 3410 & 0.44 & 1.80 & 0.01 & 2.60 \\
\enddata
\tablenotetext{a}{Calculated from stellar luminosity and effective
temperature.}
\tablenotetext{b}{Denotes a star in a binary system.}
\end{deluxetable}

\clearpage

\begin{deluxetable}{lccccccc}
\tablewidth{14.0truecm}   
\tablecaption{Correlation Coefficients and False Alarm 
Probabilities\label{corr}}
\tablehead{
   \colhead{Dataset}&
   \colhead{$r(1)$}&
   \colhead{$P_f(1)$}&
   \colhead{$r(2)$}&
   \colhead{$P_f(2)$}&
   \colhead{$r(7)$}&
   \colhead{$P_f(7)$}&
   \colhead{Slope}
}
\startdata
VBJ & 0.17 & 0.508 & 0.24 & 0.350 & 0.79 & $1.5 \times 10^{-4}$ & 
  $0.73\pm 0.14$ \\
HEG & 0.52 & 0.026 & 0.59 & $9.4 \times 10^{-3}$ & \nodata & \nodata &
  $0.88\pm 0.29$ \\
GHBC & 0.61 & 0.034 & 0.67 & 0.018 & \nodata & \nodata & $0.99\pm 0.34$ \\
CG98 & 0.11 & 0.690 & 0.16 & 0.570 & 0.82 & $1.7 \times 10^{-4}$ &
  $0.45\pm 0.10$ \\
WG01 & 0.28 & 0.225 & 0.30 & 0.198 & \nodata & \nodata & $0.45\pm 0.35$ \\
\enddata
\tablecomments{The correlation coefficient, $r$, and associated false
alarm probability, $P_f$, are given for each studies testing of the
relationships given in equations (1), (2), and (7).}
\end{deluxetable}

\clearpage

\begin{deluxetable}{lcccc}
\tablewidth{11.5truecm}   
\tablecaption{Correlating $f_{acc}$ and $R_*$ with $\dot M$ Using VBJ and 
CG98\label{closer}}
\tablehead{
   \colhead{Comparison}&
   \colhead{$r(VBJ)$}&
   \colhead{$P_f(VBJ)$}&
   \colhead{$r(CG98)$}&
   \colhead{$P_f(CG98)$}
}
\startdata
$\dot M^{1/2}$ vs. $f_{acc}$ & 0.06 & 0.82 & 0.81 & $2.5 \times 10^{-4}$ \\
$\dot M^{1/2}$ vs. $R_*^2$ & 0.50 & 0.041 & 0.28 & 0.319 \\
$\dot M^{1/2}$ vs. $f_{acc} R_*^2$ & 0.69 & $2.3 \times 10^{-3}$ & 0.84 & $8.9 
   \times 10^{-5}$ \\
\enddata
\end{deluxetable}

\clearpage

\begin{deluxetable}{lccc}
\tablewidth{8.5truecm}   
\tablecaption{Observed and Predicted Magnetic Fields\label{magp}}
\tablehead{
   \colhead{}&
   \colhead{$B_*(obs)$}&
   \colhead{$B_*(VBJ)$}&
   \colhead{$B_*(CG98)$}\\[0.2ex]
   \colhead{Star}&
   \colhead{(kG)}&
   \colhead{(kG)}&
   \colhead{(kG)}
}
\startdata
BP Tau & 2.10 & 0.76 & 4.85 \\
DF Tau & 2.30 & 1.52 & 1.22 \\
DK Tau & 2.70 & 0.98 & 4.99 \\
\enddata
\end{deluxetable}

\end{document}